\documentstyle[aps,12pt]{revtex}
\begin{document}
\title{Self-affine time series: applications and models}
\author{Jon D. Pelletier and Donald L. Turcotte}
\address{
  Department of Geological Sciences, Snee Hall, Cornell University \\
  Ithaca, NY 14853 \\
}%
\maketitle 

%
%

\noindent Table of Contents

\noindent 1. Introduction 

\noindent 2. Natural variability of climate

2.1. Temperature spectra   

2.2. River discharge and tree ring spectra

2.3. Stochastic diffusion model

2.4. Variations in solar luminosity

2.5. Drough hazard assessment

\noindent 3. Variations in sedimentation

3.1. Introduction

3.2. Surface growth model

3.3. Observations

3.4. Completeness of the sedimentary record

3.5. Bed thicknesses

\noindent 4. Variability of the earth's magnetic field

4.1. Variations of the dipole moment

4.2. Reversal record

4.3. Inclination and declination data

4.4. Model for geomagnetic variations

\noindent 5. Other applications  

\section{Introduction}
In the previous paper 
[{\it Malamud and Turcotte}, 1997, hereafter referred to as
MT] the authors considered various theoretical aspects of self-affine time series.
Several techniques for generating synthetic, self-affine time series
were discussed and alternative methods for analyzing time series 
were presented. A primary means of identifying a self-affine time series
is in terms of its Fourier spectrum. For self-affinity the power spectrum
or power-spectral density,
$S$, must have a power-law dependence on frequency, $f$: $S\propto f^{-\beta}$
(MT equation 2.4). 

When considering self-similar processes in nature there are generally upper and  
lower limits to the validity of power-law statistics. An example is a power-law
distribution for the frequency-size distribution of fragments. There will always 
be a largest and a smallest fragment. In many cases power-law statistics will
be applicable over a limited range of sizes. In other cases two power-law regimes
are found with different slopes.
There are also upper and lower limits to the self-affine behavior of
naturally occuring time series. In some cases two or more power-law regimes
are found with different values of $\beta$.   
In this paper we consider several applications of self-affine
time series in geophysics. The coverage of topics is not meant to be
complete. Instead, we consider three examples in some detail and
present applicable models.

The first application we consider is time-series data for local
atmospheric temperature. 
The spectral behavior for time scales between 200 kyr and 500 yr is
obtained from Deuterium concentrations in the Vostok ice core. Historical
temperature records are analyzed to give the spectral behavior
between time scales of 100 yr and 1 day.
The obvious daily and annual periodicities 
are removed and we focus on the stochastic content
of the time series. We find that self-affine behavior is applicable over
well-defined frequency bands. 
The self-affine behavior
is associated with interactions between the atmosphere, the space above (through
the radiation of heat),
and the oceans and continents below.      
Solutions to a stochastic diffusion equation for a layer with a substrate
reproduce the observed statistics. The results are virtually identical
to those for heat diffusion through a metallic film over a substrate and
to variations of solar luminosity.
We have also carried out power-spectral analyses of mean river discharges and
tree ring thicknesses. Both exhibit self-affine behavior with $\beta\approx
0.5$. This corresponds to a Hurst exponent $Hu\approx 0.7$, consistent with the
previous results as discussed by MT (Section 5.2). The implications of
a self-affine river discharge time series for drought assessment is also 
considered.   

In our second application we consider porosity variations 
in sedimentary basins. A model developed for the growth of atomic surface
layers is modified so that it is applicable to the spatial and temporal
variations in deposition and erosion. Self-affine variability is found
with $\beta=2$ in space and $\beta=1.5$ in time. The spatial variability
is a Brownian walk. This has been widely observed as the spectral behavior
of topography. We show that this variability is also consistent with the
spatial distribution of oil pools in sedimentary basins. The temporal
variability of sedimentation is associated with the vertical variability
of porosity. Self-affine spectra with $\beta\approx 1.5$ are good approximations
to observed data.       
The vertical variability of sedimentation and erosion can also be used to model
the completeness of the sedimentary record. It has been observed that 
the rate of sedimentation, $R$, has a power-law dependence on the time period
of sedimentation, $T$, with $R\propto T^{-0.76}$. A self-affine spectrum
with $\beta=1.5$ gives $R\propto T^{-0.75}$.

Our third application considers the variability of the earth's magnetic field.
We argue that intensity variations and
reversals of the magnetic field are a natural consequence 
of the inherent variability generated by dynamo action and magnetic diffusion
in the core.   
The field exhibits a binormal behavior and when a fluctuation crosses the 
zero intensity value a reversal occurs. The spectral behavior of the field
on time scales of 100 yr to 4 Myr has been obtained from paleomagnetic data
Over this range it is well approximated by a $1/f$ ($\beta=1$) self-affine
time series. 
Synthetic $1/f$ time series have been used to generate reversal statistics
and these are found to be in good agreement with observations. The reversal
statistics are sensitive to the values of $\beta$ and we conclude that the
agreement is strong support for $1/f$ behavior over the entire record of reversals.
A model that generates the observed $1/f$ behavior is a two-dimensional stochastic
diffusion equation.    

\section{Natural variability of climate}
\subsection{Temperature spectra}
Understanding the natural variability of climate is one of the most
important tasks facing climatologists. The {\it Intergovernmental
Panel on Climate Change} [1995] concluded that the ``balance of evidence
suggests a discernible human impact on the climate system.'' This conclusion
is based, however, on comparisons with the variability exhibited by 
general circulation models (GCM). Model 
runs often exhibit significantly lower variability, by a factor of two to five,
and a different frequency dependence on time than paleoclimatic data
[{\it Intergovernmental Panel on Climate Change}, 1995]. Other model results
give natural variability comparable in magnitude to that observed in the
last 100 years [{\it Barnett, et al.} 1992].

In this section we consider the power spectrum of temporal variations
in atmospheric temperature on time scales of 200 kyr to 1 day.
We will show that at frequencies smaller than $f\approx$ 1/(40 kyr) the power
spectrum is flat (white noise). At frequencies between $f\approx$ 1/(40 kyr)
and $f\approx$ 1/(2 kyr) the power spectrum is proportional to $f^{-2}$
(a Brownian walk). At frequencies greater than 
$f\approx$ 1/(2 kyr) the power spectrum
is proportional to $f^{-\frac{1}{2}}$. At very high frequencies
(above $f\approx$ 1/(1 month)) the spectrum varies as $f^{-\frac{3}{2}}$ 
for continental stations and remains proportional to $f^{-\frac{1}{2}}$
for maritime stations. Thus we find a sequence of self-affine spectra,
each with a characteristic values of $\beta$, over different frequency bands.

We will further show that the observed power spectrum of 
atmospheric temperature is identical to the power spectrum of variations
due to the stochastic diffusion of heat in a metallic film that is in thermal 
equilibrium with a substrate [{\it Van Vliet et al.}, 1980].
Temperature variations in the
film and substrate occur as a result of fluctuations in the heat transport
by
electrons undergoing Brownian motion. The top of the film absorbs and emits
blackbody radiation. In our analogy we associate the atmosphere with
the metallic film and the oceans with 
the substrate. Turbulent eddies in the atmosphere and oceans
are analagous to the electrons undergoing Brownian motion in a metallic
film in contact with a substrate.
     
We first consider the spectral behavior of the Deuterium concentrations
in the Vostok (East Antarctica) ice core. A 220 kyr record of temperature
fluctuations is obtained using the conversion 5.6 $\delta$D(\%)=1$^{o}$K
[{\it Jouzel et al.}, 1987]. The plot of variations in temperature versus
age is given in Figure 2.1. 
{\it Jouzel and Merlinvat} [1983] have concluded that the Vostok Deuterium record  
is a proxy for local atmospheric temperature.
Because the data are unevenly sampled we utilized the Lomb Periodogram
[{\it Press et al.}, 1992] to estimate the power spectrum. 
The results are given in Figure 2.2. We
associate the power spectrum with three regions of different self-affine 
behavior.
The first region, at frequencies less than $f\approx$ 1/(40 kyr),
is a white noise ($\beta\approx$ 0). The second region, between
$f\approx$ 1/(40 kyr) and $f\approx$ 1/(2 kyr) is a Brownian walk
($\beta\approx$ 2). In the third region, with frequencies greater
than $f\approx$ 1/(2 kyr), there is a change to a lower value of $\beta$.  
This change 
is associated with rapid variations in the Vostok core.
This is also observed
in ice cores from Greenland [{\it Yiou et al.}, 1995].
Details of this analysis have been given by {\it Pelletier} [1997a].

In order to extend our analyses to higher frequencies we have carried
out power-spectral analyses on data for atmospheric temperature variations
from weather stations. One of the longest available records is for the
average monthly temperature in Central
England, 1659-1973. The data is tabulated in {\it Manley} [1974]. 
The yearly periodicity was removed from this data by subtracting from
each value the average temperature of that month for the entire record.
The resulting time series is
given in Figure 2.3. The time series exhibits rapid fluctuations from
year to year superimposed on more gradual, lower frequency variations. 
The power spectrum estimated as the square of the coefficients of the Fast   
Fourier Transform (FFT) is presented in Figure 2.4 along with a least-square
power-law fit to the data with $\beta=$ -0.47.
We have also determined the 
average power spectrum of the time series of monthly mean temperatures
from 94 stations worldwide with the yearly trend removed.
We obtained the power spectra $S(f)$ of all
complete temperature series of length greater than or equal to 1024 months
from the
climatological database compiled by {\it Vose et al.} [1992].
The yearly trend was removed by subtracting from each monthly
data point the average temperature for that month in the 86 year record for
each station. All of the power spectra were then averaged at equal
frequency values. The results are given in Figure 2.5. 
The data yield a straight-line 
on a log-log plot with slope close to -0.5 indicating that
$S(f)\propto f^{-\frac{1}{2}}$ in this frequency range.
 
Finally we consider
the average 
power spectrum of time series of
daily mean temperature (estimated by taking the average of the maximum and
minimum temperature of each day)
from 50 continental and 50 maritime stations
over 4096 days.
Maritime stations are sites on small islands far from any
large land masses. Continental stations are well inland on
large continents, far from any large bodies of water.
We chose 50 stations at random from the
complete records (those with
greater than 4096 nearly consecutive days of data) provided
by the Global Daily Summary database compiled by the {\it National Climatic 
Data Center} [1994]. Once again the yearly periodicities were removed.
The results are given in Figure 2.6 and 2.7. Continental stations (Figure 2.6) correlate with a 
$f^{-\frac{3}{2}}$ high-frequency region.
Maritime stations (Figure 2.7) correlate with a $f^{-\frac{1}{2}}$ scaling up to the
highest frequency. The crossover frequency for the continental spectra
is $f\approx$ 1/(1 month). The difference between continental and maritime
stations results from the air mass above maritime stations exchanging heat
with both the atmosphere above and the oceans below while the air mass
above continental stations exhanges heat only with the atmosphere above it.
The three spectra have been combined in Figure 2.8 to give a 
continuous spectral behavior of local atmospheric temperature from frequencies
of 10$^{-6}$-10$^{2}$ yr$^{-1}$.  

\subsection{River discharge and tree ring spectra}
Before presenting a theoretical basis for the temperature time series
spectra given above we will consider two related time series. We first give
power-spectral analyses of hydrological time series. Figure 2.9 presents
the results of power-spectral analyses
of monthly mean river discharge data in the United States from the
Hydro-Climatic Data Network compiled by {\it Slack and Landwehr} [1994].
The annual variabilites were removed and the 
power spectra were computed in the same
manner as for the temperature data. For the streamflow data we chose
all complete records with a duration greater than or equal to 512 months and
included 636 records in our analysis. 
Since river discharges can vary by orders of magnitude between river basins
we normalized the variance of each series before averaging the spectra. 
A least-square fit to the
data a $\beta$ value of -0.50, consistent with the
value observed for the temperature data in the same frequency range. 
We have taken advantage of the
large number of available stations to investigate the possible
regional variability of the power spectra. We have averaged the power
spectra for each of the 18 hydrologic regions of the U.S. defined
by the United States Geological Survey and given in
{\it Wallis et al.} [1991]. All of the regions exhibit
the same spectral dependence with an average $\beta$ value of $-0.52$ and
a standard deviation of $0.03$, indicating little variation.
  
The second related time series we consider is the sequence of
annual tree ring widths.
We have performed spectral analyses of tree ring width chronologies
in the western
United States obtained from the International Tree Ring Database.
Tree rings in the western U.S. are strongly
correlated with precipitation
[{\it Landwehr and Matalas}, 1986].
The chronologies are time series in which the
nonstationarities in growth rates have been removed and spatial averaging
has been performed in an attempt to isolate climatic effects. Tree
ring series have the advantage of being much longer than most historical
records. We obtained 43 chronologies in the western U.S. greater than
1024 years in length. The average normalized power spectrum of those records
is presented in Figure 2.10.
The least-square fit indicates
that for tree ring time series, $S(f)$ is nearly proportional to
$f^{-\frac{1}{2}}$.

In the frequency range $f\approx$ 1/(2 kyr) to 1/(1 month) the three
data sets atmospheric temperature, river discharge, and tree
ring widths all yield spectra with a slope $\beta\approx 1/2$. In  
Section 5.2 of MT the application of the rescaled-range technique was
discussed. It was pointed out that {\it Hurst et al.} [1965]
applied the rescaled-range method to time-series data for
atmospheric temperature, river discharge, precipitation, tree
ring widths, and other climatological time series. Good correlations
were obtained with the Hurst relation (MT equation 5.5) taking Hu=0.73 on average.
From the correlation between Hu and $\beta$ given for fractional
Gaussian noises in MT Figure 5.3 we see that Hu=0.73 is entirely consistent
with the observed value $\beta\approx 1/2$.

\subsection{Stochastic diffusion model}
To see how time series with power-law power spectra
arise, we present the
results from the simulation of a
discrete,
one-dimensional stochastic diffusion process.
A discrete version of the diffusion equation for the density of
particles on a one-dimensional grid of points is
\begin{equation}
n_{i}(t_{j+1})-n_{i}(t_{j})\propto n_{i+1}(t_{j})-2n_{i}(t_{j})+n_{i-1}(t_{j})
\label{discrete}
\end{equation}
We establish a one-dimensional lattice of 32 sites
with periodic boundary conditions at the ends of the lattice.
At the beginning, we place 10
particles on each site of the lattice. At each timestep, a particle is
chosen at random and moved to the left with probability $\frac{1}{2}$ and
to the right if it does not move to the left. In this way, the
average rate at which particles leave
a site is proportional
to the number of particles in the site. The average rate at which particles
enter a site $i$ is proportional to the
number of particles on each side multiplied by one-half since the particles
to the left and right of site $i$ move into site $i$ only half of the time.
This is a stochastic model satisfying equation 2.1. 
The probabilistic nature of this model
causes fluctuations to occur in the local density of random walkers. These
fluctuations do not occur in a deterministic model of diffusion.
     
In Figure 2.7 we present the average of 50 power spectra, each spectrum
from a 
time series of the number of particles in the central site of the
32-site lattice.
The figure shows a power spectrum of the form $S(f)\propto f^{-\frac{1}{2}}$.
In Figure 2.8 
we plot the cumulative probability distribution of the time
series produced by the stochastic diffusion model. The solid circles
represent data. The curve represents the cumulative log-normal distribution
fit to the data. A good fit is obtained.

Since the distribution of values in a hydrological time series is
often log-normal,
we have shown that a simple model of stochastic diffusion gives rise
to both the power spectrum and the distribution observed for hydrological
time series.

A stochastic diffusion process can be studied analytically by adding a noise
term to the flux of a deterministic diffusion equation
[{\it Van Kampen}, 1981]:
\begin{equation}
\rho c \frac{\partial \Delta T}{\partial t}=-\frac{\partial J}{\partial x}
\label{eqthree}
\end{equation}
\begin{equation}
J=-\sigma\frac{\partial \Delta T}{\partial x} + \eta(x,t)
\label{hi}
\end{equation}
where $J$ is the heat flux,
$\Delta T$ is the fluctuation in temperature from equilibrium,
$\rho$ is the density, $c$ is the heat capacity per unit mass, $\sigma$
is the conductivity, and $\eta$ is Gaussian random noise in space and time.

Equation 2.2 is conservation of energy. Equation 2.3 is Fourier's law of
heat transport with random advection of heat superimposed. The random
advection term models the effects of local convective instabilities which
randomly advect heat vertically in the atmosphere.
{\it Novikov} 
[1963] has proposed this method for studying turbulent fluctuations.

We now determine the behavior of the stochastic diffusion model in terms of 
the power spectrum of temperature fluctuations
in a layer of width
$2l$ of an infinite, one-dimensional, homogeneous space. The
presentation we give is similar to that of {\it Voss and Clarke} [1976].
The variations in total heat energy
in the layer of width
$2l$ is determined by the heat flow across the boundaries.
Figure 2.9a illustrates the geometry of the layer exchanging
thermal energy with diffusing regions above and below it.
A diffusion process has a
frequency-dependent correlation length
$\lambda=(2D/f)^{\frac{1}{2}}$ [{\it Voss and Clarke}, 1976].
Two different situations arise as a consequence of the length scale, $2l$,
of the geometry. For high frequencies $\lambda <<2l$ and
the fluctuations in heat flow across the two boundaries are independent.
For low frequencies $\lambda >>2l$ and the fluctuations in heat across the
two boundaries are in phase. 

First we consider high frequencies. Since the boundaries fluctuate
independently, we can consider the flow across one boundary only.
The flux of heat energy is given by equation 2.2.
Its Fourier transform is given by
\begin{equation}
J(k,\omega)=\frac{i\omega\eta(k,\omega)}{\alpha k^{2}+i\omega}
\end{equation}
where $\alpha =\sigma/(\rho c)$ is the diffusivity.
The flux of heat energy out of the layer at the boundary at $x=l$ (the
other boundary is located at $x=-l$)
is the rate of change of the
total energy in the layer
$E(t)$: $\frac{dE(t)}{dt}=J(l,t)$. The Fourier transform of
this equation is
\begin{equation}
E(\omega)=-\frac{i}{(2\pi)^{\frac{1}{2}}\omega}\int_{-\infty}^{\infty}
dk\ e^{ikl}J(k,\omega)
\end{equation}
Therefore, the power spectrum of variations in $E(t)$,
$S_{E}(\omega)=<|E(\omega)|^{2}>$, is
\begin{equation}
S_{E}(\omega)\propto\int_{-\infty}^{\infty}\frac{dk}{\alpha^{2}k^{4}+\omega^{2}}
\propto\omega^{-\frac{3}{2}}
\end{equation}
In the above expression, the noise term $\eta$ does not appear because,
since it is white noise in space and time, it's average amplitude is
independent of $\omega$ and $k$, i.e. it is just a constant. 
Since $\Delta T\propto\Delta E$ the power spectrum of temperature has the
same form as $S_{E}$ and $S_{T}(\omega)\propto\omega^{-\frac{3}{2}}$.

If we include the heat flux out of both boundaries, the rate of change of
energy in the layer will be given by the difference in heat flux:
$\frac{dE(t)}{dt}=J(l,t)-J(-l,t)$. The Fourier transform of
$E(t)$ is now
\begin{equation}
E(\omega)=\frac{1}{(2\pi)^{\frac{1}{2}}\omega}\int_{-\infty}^{\infty}
dk\ sin(kl)J(k,\omega)
\end{equation}
Then,
\begin{eqnarray}
S_{T}(\omega)\propto
S_{E}(\omega)\propto\int_{-\infty}^{\infty}\frac{dk\ sin^{2}(kl)}
{\alpha^{2}k^{4}+\omega^{2}}
\nonumber\\ \propto\omega^{-\frac{3}{2}}(1-e^{-\theta}(\sin\theta
+\cos\theta))
\end{eqnarray}
where $\theta =(\omega/\omega_{o})^{\frac{1}{2}}$ and
$\omega_{o}=\alpha/2l^{2}$ is the frequency where the correlation length is equal
to the width of the layer.
When $\lambda <<2l$, the above expression reduces to $S_{T}(f)\propto
f^{-\frac{3}{2}}$. When $\lambda >>2l$, $S_{T}(f)\propto
f^{-\frac{1}{2}}$ [{\it Voss and Clarke}, 1976]. In this limit the
boundaries fluctuate in phase and heat that enters into the region from
one boundary can diffuse out of the other boundary. The result is a sequence
of fluctuations which are less persistent ($\beta$ is smaller) than the single
boundary $f^{-\frac{3}{2}}$
case. 
 
In Section 2.1 we presented evidence that
continental stations exhibit a $f^{-\frac{3}{2}}$ high-frequency region
and maritime stations exhibit $f^{-\frac{1}{2}}$ scaling up to the
highest frequency considered. This observation can be interpreted in terms of the
diffusion model presented above. The power spectrum of temperature
variations in an air mass exchanging heat by one-dimensional
stochastic diffusion is proportional to
$f^{-\frac{1}{2}}$ if the air mass is bounded by two diffusing regions
and is propotional to $f^{-\frac{3}{2}}$ if it interacts
with only one.
The boundary conditions appropriate to maritime and
continental
stations 
are presented in Figure 2.13b and 2.13c, respectively.
The maritime stations
have a $f^{-\frac{1}{2}}$ power spectrum up to the highest frequency
because the air mass above a maritime station exchanges heat with both
the atmosphere above and the oceans below. In maritime stations,
fluctuations in heat energy are readily absorbed by both the oceans
below and the atmosphere above which can radiate heat into space.
The fluctuation
calculation appropriate for maritime stations is one in which the coherent
fluctuations from two boundaries are considered as in the calculation
of the $f^{-\frac{1}{2}}$ spectrum.
The air mass above continental
stations exchanges heat energy only with the atmosphere
above it. The calculation appropriate for continental stations is the
one-boundary
model which predicts the observed $f^{-\frac{3}{2}}$ spectrum.
At low frequencies, horizontal heat exchange between continental
and maritime air masses limits the variance
of the continental stations. This crossover should occur at the time 
scale when the air masses above continents and oceans become mixed.
The time scale for one complete Hadley or Walker circulation which
mixes the air masses 
is approximately 1 month, the same time scale as the observed
crossover [{\it Pelletier}, 1997a]. 

Next we consider the stochastic diffusion model in a geometry appropriate
for a coupled atmosphere-ocean
model with an atmosphere of uniform density (equal to the
density at sea level) in thermal contact with oceans of
uniform density. The height of our model atmosphere is the scale
height of the atmosphere
(height at which the pressure falls by a factor of $e$ from its
value at sea level). Figure 2.14 illustrates the geometry
and constants chosen with $\sigma$ the vertical heat conductivity, $\rho$
the density, $c$ the specific heat per
unit mass, $\alpha$ the vertical
thermal diffusivity, and $g$ the thermal conductance of heat out of
the Earth by emission
of radiation. Primed constants denote values for the oceans.
The physical constants which enter the model are 
the density, specific heat,
vertical thermal diffusivity,
depths of the oceans and atmosphere, and the
thermal conductance by emission of radiation. The density and
specific heat of air and water are well-known constants. We chose an ocean
depth of 4\ km and an atmospheric height equal to the scale height of 8\ km as
used by {\it Hoffert et al.} [1980] in their climate modeling studies.
The eddy diffusivity we employ for the oceans is $6$x$10^{-5}$ m$^{2}$/s.
This value has been obtained from Tritium dispersion studies [{\it Garrett},
1984]. The vertical eddy diffusivity for the atmosphere we use is
1 m$^2$/s,
as 
quoted by {\it Pleune} [1990]
and {\it Seinfeld} [1986] for stable air
conditions.
This eddy diffusivity implies an equilibration time of the tropospheric
air column to be 2 years. This value is roughly consistent with the one year decay
time of the Pinatubo and El Chichon aerosols
[{\it Hofmann and Rosen}, 1987; {\it Rosen et al.}, 1994].

Since the time scales of horizontal diffusion in the atmosphere
and oceans are so much smaller than the time scales of vertical diffusion,
the rate-limiting step for thermal equilibration is vertical transport.
For this reason, we consider only the variations in local temperature 
resulting from heat exchange vertically in the atmosphere and oceans.

The equation for temperature fluctuations in space and time in the model
from equations 2.2 to 2.5 is
\begin{equation}
\frac{\partial \Delta T(x,t)}{\partial t}-\alpha (x)
\frac{\partial ^{2} \Delta T(x,t)}
{\partial x^{2}}
=-\frac{\partial\eta(x,t)}{\partial x}
\end{equation}
The mean value of $\eta$ is zero and the flux of heat of proportional to the
temperature:  
\begin{equation}
\langle \eta(x,t)\rangle=0
\end{equation}
\begin{equation}
\langle \eta(x,t)\eta(x',t')\rangle\propto\sigma(x)\langle T(x)
\rangle^{2}\delta(x-x')\delta(t-t')
\end{equation}
The delta functions indicate that the white noise term $\eta$ is uncorrelated
in space and time.
 
{\it North and Cahalan} [1981] analyzed a similar model of climate
change with respect to
predictability. They studied the diffusion equation in two dimensions as
a model for heat transport horizontally in the atmosphere. They included
a white noise term on the right side of the diffusion equation
(they used $\eta(x,t)$ where we
use $\frac{\partial\eta(x,t)}{\partial x}$) to represent variations in
heat transport by turbulent eddies. However,
a noise
term in the {\it flux} of temperature rather than in the temperature itself
is more appropriate as a model for variations in turbulent heat transfer.
 
The boundary conditions are that no heat flows out of
the bottoms of the oceans and continuity of temperature
and heat flux at the atmosphere-ocean boundary:
\begin{equation}
\sigma '
\frac{\partial T}{\partial x}|_{x=w_{2}} = 0
\end{equation}
\begin{equation}
\Delta T(x=w_{1}^{+})=\Delta T(x=w_{1}^{-})
\end{equation}
\begin{equation}
\sigma \frac{\partial \Delta T}{\partial x}|_{x=w_{1}^{-}}=
\sigma ' \frac{\partial \Delta T}{\partial x}|_{x=w_{1}^{+}}
\end{equation}
 
At the top of the atmosphere we impose a blackbody radiation boundary
condition.
Most $(65\%)$ of the energy incident on the Earth is reradiated as long-wavelength
blackbody radiation from the H$_{2}$O and CO$_{2}$ in the atmosphere
[{\it Peixoto and Oort}, 1992]. This
radiated energy depends on the temperature of the
atmosphere at the point of emission according to the Stefan-Boltzmann
law. It is common practice to assume that temperature
variations about equilibrium are small. This is a good approximation
since the global mean temperature has
fluctuated by only about ten degrees Kelvin during the last glaciation as
illustrated in Figure 2.1.
With a linear approximation, the radiated energy will be
proportional to the temperature difference from equilibrium [{\it Ghil}, 1983].
The boundary
condition at the scale height
of the atmosphere (which we take to be
representative of the average elevation where radiation is emitted
from the atmosphere) is then
\begin{equation}
\sigma \frac{\partial \Delta T}{\partial x}|_{x=0}=g\Delta T(x=0)
\end{equation}
We will use the value $g=1.7\ W/m^{2}K$ as used by {\it Ghil} [1983] and
{\it Harvey and Schneider} [1985]. It is often assumed that a feedback
exists between atmospheric or sea-surface temperature and cloud cover.
If such a feedback existed, it might be necessary to parameterize radiated energy 
in terms of cloud cover or atmospheric water vapor. However, 
no evidence for such a feedback has been
found [{\it Arking and Ziskin}, 1994].

The existence of two layers of different diffusivity makes the study of
the two-layer model much more complex than for the one-layer models applied
to the atmosphere above the continents and the oceans.
{\it Van Vliet et al.} [1980] used Green's functions to solve this two-layer
model.
The Green's function of the Laplace-transformed
diffusion equation is defined by
\begin{equation}
i\omega G(x,x',i\omega)-\alpha (x)\frac{\partial^{2}G(x,x',i\omega)}
{\partial x^{2}}=\delta (x-x')
\end{equation}
where G is governed by the same boundary conditions as $\Delta T$.
This equation can be solved by separating $G$ into two parts: $G_{a}$ and
$G_{b}$ with $x < x'$ and $x > x'$, respectively. $G_{a}$ and
$G_{b}$ satisfy the homogeneous (unforced) diffusion equation with a
jump condition relating $G_{a}$ and $G_{b}$:
\begin{equation}
\frac{\partial G_{a}}{\partial x}|_{x=x'}-\frac{\partial G_{b}}{\partial x}
|_{x=x'}=\frac{1}{\alpha (x')}
\end{equation}
 
The power spectrum
of the average temperature in the atmosphere
in terms of G is given by {\it Van Vliet et al.} [1980] as:
\begin{equation}
S_{\Delta T}(f)\propto
Re (\int_{0}^{w_{1}}\int_{0}^{w_{1}} G_{1}(x,x',i\omega)
dx dx')
\end{equation}
\begin{eqnarray}
\propto
Re(\int_{0}^{w_{1}}\int_{0}^{x}G_{1b}(x,x',i\omega)dx dx' \nonumber\\ +
\int_{0}^{w_{1}}\int_{x}^{w_{1}}G_{1a}(x,x',i\omega)dx dx')
\end{eqnarray}
where $G_{1}$ stands for the solution
to the differential equation for G where the source point is located in
the atmosphere and $Re$ denotes the real part of the complex expression.
Two forms of $G_{1a}$ and $G_{1b}$ are necessary for $x$
located above and below $x'$, respectively, due to the discontinuity
in the derivative of $G_{1}$ created by the delta function.
The solution of $G_{1}$ which satisfies the above differential equation and
boundary conditions is
\begin{eqnarray}
G_{1a}=\frac{L}{\alpha K}(\frac{\sigma 'L}{\sigma L'}\sinh(\frac{w_{1}-x'}{L})
\sinh(\frac{w_{2}}{L'})\nonumber\\ +\cosh(\frac{w_{1}-x'}{L})
\cosh(\frac{w_{2}}{L'}))
(\sinh(\frac{x}{L})+\frac{\sigma}{Lg}\cosh(\frac{x}{L}))
\end{eqnarray}
and
\begin{equation}
G_{1b}=G_{1a}+\frac{L}{\alpha}\sinh(\frac{x'-x}{L})
\end{equation}
where
\begin{eqnarray}
K=(\sinh(\frac{w_{1}}{L})+\frac{\sigma}{Lg}\cosh(\frac{w_{1}}{L}))
\frac{\sigma 'L}{\sigma L'}\sinh(\frac{w_{2}}{L'})
\nonumber\\ +(\cosh(\frac{w_{1}}{L})
+\frac{\sigma}{Lg}\sinh(\frac{w_{1}}{L}))\cosh(\frac{w_{2}}{L'})
\end{eqnarray}
and $L=(\alpha/i\omega)^{\frac{1}{2}}$ and
$L'=(\alpha '/i\omega)^{\frac{1}{2}}$. Performing the integration
{\it van Vliet et al.} [1981] obtained
\begin{eqnarray}
S_{\Delta T}(f)\propto
Re (L^{2}(\frac{\sigma 'L}{\sigma L'}
\tanh(\frac{w_{2}}{L})((\frac{gw_{1}}{\sigma}-1)
\nonumber\\
\tanh(\frac{w_{1}}{L})
-\frac{2gL}{\sigma}\frac{\cosh(w_{1}/L)-1}{\cosh(w_{1}/L)}+\frac{w_{2}}{L})
\nonumber\\
+(\frac{gw_{1}}
{\sigma}+(\frac{w_{1}}{L}-\frac{gL}{\sigma}\tanh(\frac{w_{1}}{L}))
((\tanh(\frac{w_{1}}{L})+\frac{\sigma L}{g})
\nonumber\\ \frac{\sigma 'L}{\sigma L'}
\tanh(\frac{w_{2}}{L'})+(1+\frac{\sigma}{Lg}
\tanh(\frac{w_{1}}{L})))^{-1}
\label{bigone}
\end{eqnarray}
For very low frequencies several approximations can be made: 
\begin{equation}
\tanh(\frac{w_{1}}{L})\approx\frac{w_{1}}{L},
\tanh(\frac{w_{2}}{L'})\approx\frac{w_{2}}{L'}
\end{equation}
\begin{equation}
\frac{\cosh(w_{1}/L)-1}{\cosh(w_{1}/L)})\approx\frac{1}{2}
\frac{w_{1}^{2}}{L^{2}}
\end{equation}
Reducing equation 2-26,
\begin{equation}
S_{\Delta T}(f)\propto \frac{1}{1+\omega^{2}/\omega_{0}^{2}}
\propto \frac{1}{f^{2}+f_{0}^{2}}
\end{equation}
This is the low-frequency Lorentzian spectrum observed in the Vostok data.
The crossover frequency as a function of the constants chosen
for the model is
\begin{eqnarray}
f_{0}=\frac{g}{w_{1}c\rho + w_{2}c'\rho '(1+g w_{1}/\sigma)}
\end{eqnarray}

At higher frequencies the following approximations hold
\begin{equation}
\tanh(\frac{w_{1}}{L})\approx\frac{w_{1}}{L},
\tanh(\frac{w_{2}}{L'})\approx 1
\end{equation}
\begin{equation}
\frac{\cosh(w_{1}/L)-1}{\cosh(w_{1}/L)}\approx\frac{1}{2}\frac{w_{1}^{2}}{L^{2}}
\end{equation}
then
\begin{equation}
S_{T_{av}}(f)\propto \frac{1}{2}(\frac{2gw_{1}}{\sigma})^{\frac{1}{2}}
(\frac{c\rho\sigma}{c'\rho '\sigma '})^{\frac{1}{2}}
(\frac{g}{w_{1}\rho c f})^{\frac{1}{2}}\propto f^{-\frac{1}{2}}
\end{equation}
This is the broad $f^{-\frac{1}{2}}$ region observed in
the power spectrum of the 
temperature data and predicted based on the simpler one-layer model 
exchanging heat with regions above and below.
The high- and low-frequency spectra meet at
\begin{equation}
f_{1}=\frac{g}{w_{1}\rho c}(\frac{\sigma}{2gw_{1}})^{\frac{1}{3}}
(\frac{c'\rho'\sigma'}{c\rho\sigma})^{\frac{1}{3}}
4^{\frac{1}{3}}(\frac{c\rho w_{1}}
{c'\rho ' w_{2}})^{\frac{4}{3}}
\end{equation}
\begin{equation}
\approx 1/(10 kyr)
\end{equation}
This value agrees within an order of magnitude
to that observed in the Vostok data ($f\approx$
1/(2 kyr)).

At time scales less than 2 kyr, fluctuations in heat energy input 
into air masses from the 
oceans or variations in emitted radiation can be absorbed by the other reservoir.
At
lower frequencies, the atmosphere and oceans are in thermal 
equilibrium. The oceans can no longer absorb thermal fluctuations input into the
atmosphere from fluctuations in radiative emission at this time scale and vice versa.
The variance in temperature of the atmosphere and oceans 
is then determined solely by the radiation boundary condition. The fluctuating
temperature at the top of the atmosphere will result in a white noise 
flux out of the atmosphere-ocean system. The average temperature of the 
atmosphere and oceans will be given by the sum of a white noise, a Brownian
walk. This is observed in the Vostok data. The Brownian walk behavior of
the climate system flattens out at low frequencies as a result of
a negative feedback mechanism: as the coupled atmosphere and oceans warm up
(or cool down) due to nonstationary fluctuations of the random
heat exchange from radiative emission, the system will 
radiate, on average, more (less) radiation, limiting the variance at low
frequencies. This interpretation depends on the
conductivity being low compared to the 
time scale for radiative damping, $\tau =c'\rho ' w_{2}/g$. If so, the rate-limiting 
step of radiative equilibrium is the conduction of heat from the oceans through the
atmosphere. The time constant of this thermal discharge (analagous to the 
electrical discharge of an RC circuit) is then
$\tau =\sigma/w_{1}w_{2}c'\rho '$.
However, if the conductivity
is high,
the time scale of radiative equilibrium is determined by the time scale for
radiative damping, $\tau =c'\rho ' w_{2}/g$. Using the estimates
for the thermal diffusion properties of the atmosphere and the oceans listed
in Figure 2.14,
the conductivity is large and the low-frequency portion of the spectrum is 
governed by the radiative damping constant. This is problematic, however, 
because the time scale of radiative equilibrium is then 
$\tau =c'\rho ' w_{2}/g=$ 600 yr. Such a short time scale implies that
no $f^{-2}$ spectral region should exist at all. An atmospheric 
vertical diffusivity of 0.1 m$^{2}$/s is required in order for the model
spectrum to be consistent with the spectrum of the Vostok data with the 
$f\approx$ 1/(40 kyr) crossover frequency. This is an order of magnitude
smaller than the values quoted by {\it Pleune} [1990] and {\it Seinfeld} [1986]. 

Besides the frequency dependence of the power spectrum, the model we
have presented predicts that the distribution of temperature variations
from equilibrium obeys a Gaussian distribution. This is because the stochastic
term obeys a Gaussian distribution function and the temperature
fluctuations are related to the stochastic term through a linear transformation.
By definition, the probability density function is only defined for time scales
in which
the temperature fluctuation time series are stationary. Gaussian time series
with power-law power spectra of the form $S(f)\propto f^{-\beta}$ are stationary
if $\beta < 1$ 
and non-stationary if $\beta\geq 1$ (MT section 5). 
Thus, a unique probability
density function 
only exists for very long time scales (greater than 100 kyr)
where the power spectrum is constant ($\beta =0$) and for the range of time
scales in which the power spectrum obeys $S(f)\propto f^{-\beta}$ with
$\beta=\frac{1}{2}$. {\it Matteucci} [1990] has computed the probability distribution
function for climatic variations at very long time scales with the SPECMAP
stack. He obtained a Gaussian distribution. Similarly, {\it Janosi and Vattay} [1992]
have obtained a Gaussian distribution with monthly temperature datasets of several
decades length with the
annual variability removed.

{\it Manabe and Stouffer} [1996] have completed power-spectral analyses of
variations in local atmospheric temperature in control runs of a
coupled atmosphere-ocean-land surface model. They computed the power
spectrum of temperature time series of each surface grid point and
then averaged the power spectra at equal frequency values, as in our
observational power-spectral analyses. Their results are presented in
Figure 2.15. They found different spectra for
continental and maritime gridpoints. Maritime gridpoints exhibited
power-law power spectra from time scales of one month to several hundred
years with an exponent of close to $-0.25$. Continental gridpoints,
however, showed flat spectra up to time scales of about 100 yr,
in contrast to observations. Exploring the similarities and differences
between the approach in this paper, GCM results such as
those of {\it Manabe and Stouffer} [1996], and observations should enable us
to learn more about this fundamental problem in Earth science.
 
Time-series analysis of paleoclimatic data often exhibit a dominant
peak near 100 kyr. Although variations in the eccentricity of the
Earth's orbit occur with this frequency, this variation is not expected
to produce a linear influence on climate change since this orbital
variation results in only a fraction of a percent change in the amount
of radiation incident on the Earth [{\it Hays et al.}, 1976]. Although there are
nonlinear models
that predict a 100 kyr periodicity, it is generally agreed that the
underlying mechanism for this peak is not well understood [{\it Kerr}, 1978].
The model
presented in this section leaves the question open as it does not predict
any periodicity. The only component of the system thought to have a
characteristic time scale of 100 kyr is the cryosphere [{\it Mitchell}, 1976].
Perhaps the cryosphere can produce a 100 kyr peak in the power spectrum
when forced by the background spectrum predicted by the model of this paper.
Studies incorporating the cryosphere into our model are an
important extension of our work that may lead to new insights into the
nature of the 100 kyr periodicity.

\subsection{Variations in solar luminosity}
We have applied the same model presented in this Section 4.4 to variations in
the solar luminosity from time scales of minutes to months [{\it Pelletier}, 1996].
In Figure 2.16 we present the power spectra estimated with the Lomb periodogram 
of ACRIM
solar irradiance data sampled during
1987 and 1985 plotted as a function of the frequency in hours$^{-1}$.
The same sequence of power-law behavior is observed in these data as
are observed in the Vostok data. Large peaks
appear at the orbital frequency of the satellite and its harmonics. These
peaks are an artifact of the spectral estimation.
A stochastic diffusion model of the turbulent heat transfer between the
granulation layer of the sun, modeled as a homogeneous thin layer with a
radiative boundary condition, and the rest of the convection zone,
modeled as a homogeneous thick layer with thermal and diffusion constants
appropriate to the lower convection zone, predicts the same spectral form
observed in solar irradiance data.
The time scales of thermal
and radiative equilibrium of the solar convection zone based upon
thermal and diffusion constants estimated from mixing-length theory
match those observed in the ACRIM data. Further details are discussed in
{\it Pelletier} [1996]. 
 
\subsection{Drought hazard assessment}
One of the principle applications of time-series analysis is to drought 
hazard assessment. A major question is whether ``short-memory'' models
are adequate or whether ``long-memory'' models such as self-affine
noises and walks are required [{\it Bras and Rodriguez-Iturbe}, 1985]. 

Since hydrologic droughts
are
phenomena requiring multiple years of low flow, the frequency
of occurence will be affected by correlations in the time series
of discharge.
We now illustrate how fractional noises
can be used to estimate drought
frequencies. The use of fractional noises that
exhibit the Hurst phenomenon has been proposed by
{\it Booy and Lye} [1989] for use in flood-frequency analysis.
The goal of stochastic
hydrology is to generate synthetic time series of river discharge that 
accurately reproduce hydrological time series. Based on 
evidence for the applicability of a fractional noise with $\beta\approx 1/2$,
we generated synthetic
time series with two-parameter log-normal distributions
that fit the historical records of river discharge.
We first discuss the
techniques and results of drought
frequency analyses for series with different log-normal distributions.
Then we discuss the results of a comparison between drought
frequencies for the Colorado river based upon a fractional noise
with exponent of $-1/2$ and a short-memory AR(1) model.

Techniques for generating synthetic log-normal fractional noises have been discussed
by MT, Section 4.2. We utilize synthetic noises with $\beta=1/2$ and
$c_{v}=$ 0.2, 0.4, and 0.6.
There is no unique definition of a drought; several alternatives 
were discussed in a recent drought assessment of the southwestern U.S. by
{\it Tarboton} [1994]. Perhaps the most straightforward definition is
that proposed by {\it Yevjevich} [1967] and {\it Dracup et al.}
[1980]. They defined
a drought as any year or consecutive number of years during which average
annual streamflow is continuously below the long-term mean annual
runoff. The magnitude is the average deficit during the drought.
The principal drawback to this definition is that two 5-year
droughts separated by one wet year will only be recognized as 5-year
droughts even though the succession of droughts results in
ten or eleven years of critically low supply.
In Figure 2.17 we present the results of drought frequency analyses
based on this definition of a drought. Each part is a
two-dimensional contour plot of the logarithm (base 10) of the
recurrence interval of a drought of a given duration and magnitude,
with the magnitude normalized to the mean flow. Figure 2.17 a, b, and c  
represent coefficients of variation 0.2, 0.4, and 0.6, respectively.
To construct each figure we generated synthetic records of one
million years in length and searched them for drought occurences.

In order to assess the importance of long-range persistence on
the likelihood of severe drought, we have completed a drought
frequency analysis using an AR(1) model 
for the Colorado river at Lees Ferry [{\it Kendall and Dracup}, 1991]
with a fractional noise
model for the same location.
The synthetic time series were one million years in length.
We found that for droughts of ten years duration and small magnitude,
100-year events according to the fractional noise model, the difference
in recurrence interval for the two models is a factor of five. We conclude
that the presence of long-range persistence has a significant effect
on the likelihood of severe drought.
The presence of long-range persistence does not, however, appear to improve
the ability to predict future climatological and hydrological time
series to any significant degree [{\it Noakes et al.}, 1988].

\section{Variations in sedimentation}
\subsection{Introduction}
\setcounter{equation}{0}
We now turn to porosity variations in sedimentary basins. We show that
these variations are self-affine walks in both the horizontal and vertical 
directions. We show that the observed distributions can be reproduced
using a standard model for surface growth. We will further show that
this model is consistent with the observed variations and
episodicity in sediment deposition.

In the past decade many studies have documented the scale-invariance 
of porosity and density variations in sedimentary basins. Power-law
power spectra of vertical density and porosity well logs have been reported by 
{\it Hewett} [1986], {\it Walden and Hosken} [1985], {\it Pilkington
and Todoeschuck} [1990], {\it Todoeschuck et al.} [1990], {\it Holliger}
[1996], {\it Shiomi et al.} [1996], {\it Dolan et al.} [1996], and {\it
Pelletier and Turcotte} [1996]. {\it Tubman and Crane} [1995] and
{\it Deshpande et al.} [1996] have presented evidence for scale-invariant
horizontal variations in density and porosity from well logs and seismic data.
In addition, {\it Dunne et al.} [1995] presented evidence that the 
topography of alluvial plains along the channel strike is also
scale-invariant. They performed spectral analyses on fluvial
microtopographic transects of an alluvial plain in Kenya. They 
obtained power spectra
with $S(k)\propto k^{-\beta}$ where $k$ is the wave number
and $\beta$ ranges from 1.5 to 2.
Based on his original observation, {\it Hewett} [1986] 
developed a fractal-based interpolation
scheme for determining the three-dimensional
porosity variations in sedimentary basins
using available well logs. The validity of the interpolated
structure was subsequently verified in a variety of ways.
This approach was applied to modeling ground water migration by {\it Molz
and Boman} [1993]. 

\subsection{Surface growth model}
Before considering the observed spectra further, we present a model for
surface growth which will be predictive of both vertical and horizontal  
porosity variations in sedimentary basins. At each
time step,
a site on a one-dimensional lattice is chosen at random.
During that time step,
a unit of sediment is deposited on that site or on one of its nearest
neighbors, depending on which site has the lowest elevation. This is the
simplest model combining randomness and the tendency for sediment to be deposited
in low-lying areas of an alluvial plain.
The model is illustrated in Figure 3.1.
The cross-hatched block shows the
unit of sediment being added to the surface. In each case, an arrow points towards
the site upon which the unit of sediment will be deposited.
In Figure 3.1a the chosen site has a lower
elevation than either of its nearest neighbors, so the sediment is deposited
at the chosen site. In Figure 3.1b one of the nearest neighboring sites has a
lower elevation and the sediment is deposited at that lower site. In the
case of a tie for the lowest elevation between
two or three sites, the site on which the sediment
is deposited is chosen randomly between the sites of the same elevation,
as in Figure 3.1c. 
The local elevation is the total number of units of sediment that have
been deposited at the site.

This model
of surface growth was first analyzed by {\it Family} [1986] with applications
to the growth of atomic surface layers. He reported the
results of computer simulations which showed that the model
produces scale-invariant variations of the surface in space and time.
He found that the standard deviation, $\sigma$,
of the surface follows the relation
\begin{equation}
\sigma(L,T)\propto L^{\frac{1}{2}}T^{\frac{1}{4}}
\end{equation}
where $L$ is a length scale and $T$ is a time scale.
Surfaces with scale-invariant standard deviations $\sigma(L,T)\propto L^{Ha_{x}}
T^{Ha_{t}}$ have a power-law dependence of the power-spectral density, $S(k)$,
on wave number $k$
of the form $S(k)\propto k^{-2Ha_{x}-1}$ (i.e.\ $\propto k^{-2}$ for
$Ha_{x}=1/2$)
and a power-law dependence on frequency
of the form
$S(f)\propto f^{-2Ha_{t}-1}$ (i.e.\ $\propto f^{-\frac{3}{2}}$ for
$Ha_{t}=1/4$).

An example of the surface elevation produced by the model with 1024
grid points is given in Figure 3.2. The average dependence of the
power-spectral density on wave number from fifty independent simulations
is given in Figure 3.3. The power spectrum is proportional to $k^{-2}$
indicating that the surface is a Brownian walk. Other lattice 
sizes yield similar results. The synthetic topography given in
Figure 3.2 is very similar to the one-dimensional transect of
Kenyan topography obtained by {\it Dunne et al.} [1995] and plotted
in Figure 3.4.
{\it Hooke and Rohrer} [1979] have mapped the 
topographic profiles of alluvial fans perpendicular to the flow direction.
The synthetic profile is also strikingly similar to their alluvial fan
profiles.

In Figure 3.5 we plot the variations in surface elevation (subtracted from the
mean height of the landscape) at the central
site of our simulation after the simulation has reached a dynamic steady state. In Figure
3.6 we 
present the average power spectrum of the difference from the mean height
of the central site produced in fifty simulations. The power spectrum is
proportional to $f^{-\frac{3}{2}}$. If
deposition and erosion took place independently, the variation of the
local elevation relative to the mean with time would
be a random walk 
with power spectrum $S(f)\propto f^{-2}$. The random-walk model
has been analyzed as a model for sedimentary 
bed formation. The effect of the diffusion
term is to preferentially fill low-lying areas of the alluvial plain.
This results in an anticorrelated sequence of deposition and erosion:
after an area has aggraded it has a higher elevation and a lower rate of
future aggradation.
Without the presence of the
diffusion term, the surface would be white noise. The random-walk model,
therefore, results in a very unrealistic alluvial plain topography.
    
We can also include the effects of erosion in our model.
Although deposition generally occurs
in topographic depressions, tending to smooth out the floodplain,
erosion is less consistent. Erosion can downcut in a channel or, during
a large flood, can lower alluvial ridges.
We have modified our simulation to include the effects of erosion by
choosing randomly at each timestep whether to deposit or erode sediment during
that timestep. The probability of deposition must be greater than 0.5
in order to accumulate a sedimentary basin over time.
We have studied the above model assuming that erosion occurs preferentially
on channel floors, randomly on the landscape, or preferentially on alluvial
ridges.
In the simulations in which we assumed erosion to occur preferentially on
the channel floors, we have included an erosion rule that takes away rather
than deposits a unit of sediment at a randomly chosen site or one of its
nearest neighbors, depending on which has the lowest elevation. We have
also investigated rules that remove a unit of sediment only from
the chosen site (to simulate
random erosion on the floodplain) and a rule that removes sediment from
the chosen site or one of its nearest neighbors, depending on which site
is highest, to simulate the preferential erosion of alluvial ridges.
The exponents of the power-law power
spectra obtained in the model without erosion are
unchanged for any of these erosion models.

In the simplest version of the model which includes only deposition the
probability that a particle is added to
the site is proportional to two if both of a site's neighbors have a higher
elevation, proportional to one if only one of the neighbors is higher, and
zero if both neighbors are lower.
The model may be described mathematically with
a stochastic difference equation of the form
\begin{equation}
h_{i,t+1}-h_{i,t}\propto \Theta (h_{i+1,t},h_{i,t})+ \Theta (h_{i-1,t},h_{i,t})
\end{equation}
where $h_{i,t+1}-h_{i,t}$ represents the most probable growth
rate of the surface
and $\Theta$ is the Heavyside
function defined by
$\Theta (x,x_{o})=1$ if $x>x_{o}$ or $0$ if $x<x_{o}$.
Averaging this equation over a time long compared to the time required to
grow
a single layer of unit height of sediment, the
equation for the average surface growth rate is
\begin{equation}
\langle h_{i,t+1}-h_{i,t}\rangle\propto\langle h_{i+1,t}-h_{i,t}\rangle
+\langle h_{i-1,t}-h_{i,t}\rangle
\end{equation}
\begin{equation}
\propto \langle h_{i+1,t}\rangle-2\langle h_{i,t}\rangle+\langle h_{i-1,t}
\rangle
\end{equation}
This is a discrete version of the diffusion equation. Directing
sediment to lower elevations smooths out the surface and is equivalent to a
diffusion process. As recognized by {\it Family} [1986],
a continuous version of the discrete model
is provided by a one-dimensional diffusion equation with a
Gaussian white noise term:
\begin{equation}
\frac{\partial h(x,t)}{\partial t}=D\frac{\partial^{2}h(x,t)}
{\partial x^2}+\eta (x,t)
\label{ew}
\end{equation}
 
The term $\eta (x,t)$ represents actual deposition and erosion. The assumed
Gaussian white noise is characterized by a mean, $\overline{\eta}$, and
a standard deviation $\sigma$. If $\overline{\eta}=0$ there is no net
deposition and sedimentation balances erosion. For $\overline{\eta}>0$ there
is net deposition and for $\overline{\eta}<0$ there is net erosion. The
ratio $\sigma/\overline{\eta}$ is a measure of the amplitude of
fluctuations in the sedimentation process. The diffusion term in equation 3.5
introduces both spatial and temporal correlations in the sedimentation process
not present in the random-walk model.

\subsection{Observations}
We will next consider some observed spectra of the vertical variations
of porosity in sedimentary basins and compare them with the results we
have obtained. 
Porosity as a function of depth is routinely measured at equal intervals
in formation well logs [{\it Hewett}, 1986].
As a specific example we have considered
porosity logs from 15 wells in the Gulf of
Mexico. One of the logs is plotted in Figure 3.7.
The wells are drilled in a deltaic sedimentary environment with
a few large, nearly vertical faults [{\it Alexander}, 1995].
The power spectra for these wells are given in Figure 3.8.
as a function of the wavenumber $k$ in
m$^{-1}$. At spatial scales larger than $\approx$ 3\ m the power spectra are well
approximated by a power law. Below
this scale the power-spectral density decreases sharply in most of the wells.
This decrease may
be the result of a transition from second-order heterogeneities
(dominated by variations in porosity within the larger genetic
units) to third order heterogeneities which result from the geometrical
arrangements of individual depositional units. The transition
from second to third-order heterogeneities
occurs at the scale of meters [{\it Allen and Allen}, 1990] and is
consistent with the 3 m scale of the break observed in
the power spectra.
We estimate $\beta$ from
the slope of the least-square linear fits to the log-log plots at scales
greater than 3 m.
The values of $\beta$ obtained exhibit
considerable variability from well to well. However, the average $\beta
=1.4$ is close to the value 1.5 predicted by the model.
The standard deviation is 0.2.

{\it Dolan et al.} [1996] 
reported ranges of values of $\beta$ from time-series analyses of
vertical density and porosity variations in well logs.
They obtained average power spectral exponents
$\beta=$ -1.50, -1.62, and -1.46 using three different numerical procedures
for a cluster of ten wells in a fluvial sedimentary environment. These values
are in excellent agreement with the ones we obtained and with our model.
{\it Holliger} [1996] has reported values of $\beta$ 
from 1.2 to 1.4, somewhat smaller,
but roughly consistent with the values reported here.

We will next consider several observational studies relevant to the
horizontal variations of porosity in sedimentary basins. Direct
measurements of topography on relevant scales have been carried out 
by {\it Dunne et al.} [1995]. These authors
have performed power spectral analyses of
fluvial microtopographic transects perpendicular to the fall line from two
hillslopes obtained
with laser altimetry from scales of 0.1 to 100\ m. 
Their work provides us with a direct test of
our model for the topographic variations of an alluvial plain.
They obtained power spectra with a power-law dependence on wave number as
predicted. The exponents of the
power spectra had an average of $\beta=$ 
1.6 with a standard deviation of 0.2, somewhat
smaller than our model prediction of $\beta=$ 2. 

In addition to the power-spectral behavior of the surface in space and time
discussed above, the stochastic diffusion model also predicts a Gaussian
distribution of the surface elevation. This is true of any linear stochastic
differential equation with Gaussian noise. The microtopographic
transects obtained by {\it Dunne et al.} [1995] enable us to
test this prediction. In Figure 3.9 we present the distribution
of elevations from the 15 profiles published by the authors. The profiles
were digitally scanned for the analysis. Also plotted in Figure 3.9
is the non-linear least-square fit to a Gaussian distribution. A good fit
is obtained.

We will next show that the distribution of producing oil and gas wells 
is consistent with $\beta$=2 horizontal porosity variations.
{\it Barton and Scholz}
[1995] have presented the spatial distribution of drilled
wells and wells showing hydrocarbons
in the  Denver and Powder River basins. These
basins evolved from sediment deposition in a
meandering alluvial environment [{\it Berg}, 1968]. Using the
box counting technique, {\it Barton and Scholz} [1995]
found that the fractal dimensions for the drilled
wells in the two basins were 1.80 and 1.86 and that the fractal dimensions
of wells showing hydrocarbons were 1.43 and 1.49, respectively.
After petroleum is generated and migrates from source rocks,
it will move from sites of high potential energy to sites of low potential energy.
Hydrocarbons are found in traps that are the crests of
low-porosity caprock that have obstructed
its upward migration [{\it Allen and Allen}, 1990]. The caprock will mimic the floodplain
relief at the time of its deposition. This is consistent with the observation that
hydrocarbons are often found in geometries which mimic the topography of the
alluvial plain at the time of deposition in a variety of fluvial depositional
environments such as meandering [{\it Curry and Curry}, 1972], deltaic [{\it Coleman
and Prior}, 1982], and submarine fans [{\it Garcia}, 1981; {\it Wilde et al.}, 1978].
A simple model for the horizontal spatial distribution of hydrocarbons in a
reservoir is one in which
hydrocarbons are assumed to be accumulated in all of the crests of the caprock
above a certain elevation. 

The spatial distribution of wells showing hydrocarbons in the Powder River and Denver
basins are given in Figure 3.10.
We have set the width of each basin
to be 128 units so as to facilitate comparisons with a synthetic reservoir
constructed on a 128 x 128 grid. We analyzed the data with the pair-correlation
function which we believe to be a better
estimator of correlations for point processes than
box counting. 
  
The two-dimensional pair-correlation function $C(r)$ is defined as the
number of pairs of wells whose separation is between $r$ and
$r$ + $\Delta r$, per unit area
[{\it Vicsek}, 1992]. The pairs are binned in logarithmically spaced
intervals $\Delta r$.
For a data set with scale invariant clustering, $C(r)\propto r^{-\alpha}$
where $\alpha$ is related to the fractal dimension through $D=2-\alpha$
in two dimensions [{\it Vicsek}, 1992]. The pair-correlation function
is commonly employed in the analysis of diffusion-limited-aggregation. However,
studies incorporating it in the earth sciences are rare. {\it Kagan and
Knopoff} [1980] have applied it to  the spatial clustering
of earthquakes.
Figure 3.11 shows the pair-correlation function of the Denver and Powder
River basin wells on a log-log plot.
The least-square
fits to the correlation function yield exponents of $\alpha=-0.59$ for
Powder River and $\alpha=-0.50$ for the Denver basin, implying
$D=1.41$ and $D=1.5$, respectively.
The results obtained by the
pair-correlation method are in close agreement with the results obtained
by {\it Barton and Scholz} [1995] using box counting.

To show that these correlation functions are consistent with a caprock with
Brownian walk topography,
we have constructed synthetic reservoirs
where hydrocarbon traps are regions where the caprock elevation is larger
than a threshold value. In order to do this
we synthesized two-dimensional fractional Brownian walks on a 128 x 128
lattice with the
Fourier-filtering technique discussed in MT Section 3.1.
The threshold value for showing hydrocarbons
was chosen such that the resulting synthetic reservoir had the same
percentage of showing wells as the Denver and Powder River basins (about 5\%).
Figure 3.12
shows a synthetic reservoir produced with $\beta =2.0$ (a Brownian walk).
The synthetic reservoir shows a degree of clustering similar to the Denver and
Powder River basins. In Figure 3.9 we have plotted the
pair-correlation
functions for the showing wells in synthetic reservoirs constructed with
$\beta =2.5,\ 2.0,\ 1.5$, and $1.0$.
The pair-correlation functions show a
gradual decrease with decreasing $\beta$. The synthetic reservoirs
whose
scaling exponents $\alpha$ most closely match those of the Denver and
Powder River basins are $\beta =2.0$ and $\beta =1.5$. Although we
cannot
precisely determine the scaling exponent of the porosity variations with this
method, we conclude that $\beta$ is close to $2$,
consistent with our model.

Besides the pair-correlation function, two other fractal relations
allow us to infer Brownian walk paleotopography from horizonal variations in
sedimentary basins.
{\it Agterberg} [1982]
has computed the fractal dimension of the perimeter of sand isopach contours
from the Lloydminster oil field to be 1.3, close to the value of  1.25
measured for coastlines
and topographic contours [{\it Turcotte}, 1992]. {\it Barton and Scholz} [1995]
have presented frequency-size distributions of oil pools. They found 
that the cumulative number of oil fields
has a power-law dependence on the volume of the fields with exponent close
to -1: $N(>V)\propto V^{-1}$. {\it Kondev and Henley} [1995] have related the
length distribution of contour lengths of Gaussian surfaces to the Hausdorff 
measure $Ha$. {\it Pelletier} [1997b]
has shown that their results imply that
the cumulative frequency-area distribution of areas enclosed by contours of a
Brownian walk surface is $N(>A)\propto A^{-3/4}$. Since oil fields have
a much larger horizontal extent than vertical extent, it is reasonable to
assume that area and volume are proportional. Our model
of the migration of hydrocarbon into
regions with caprock topography above a threshold elevation then predicts
$N(>V)\propto V^{-3/4}$ in reasonable agreement with the cumulative
frequency-size distributions of {\it Barton and Scholz} [1995].
{\it Pelletier} [1997b] has employed the same techniques to infer the
self-affinity of the top of the convective boundary layer from the
size-distribution of cumulus cloud fields. 

\subsection{Completeness of the sedimentary record}
A related problem to topography and porosity variations in sedimentary
basins is the statistics of preserved sections.
Stratigraphic sections are formed by alternating periods of deposition and erosion
or non-deposition. The resulting stratigraphic section contains the
deposited sediments that were not subsequently eroded. Various stochastic models have
been proposed to explain aspects of sedimentary bed formation, including the
frequency distribution of bed thicknesses.
Beginning with Kolmogorov's
work [{\it Kolmogorov}, 1951],
many studies have investigated random-walk models of sedimentation.
Random-walk models assume
that the magnitude
of alternating depositional and erosional events are independent.
These models are applied by letting the typical episodes of deposition and erosion
define minimal units of a discrete time scale. The lengthy periods of non-deposition,
as well as any long intervals of deposition and erosion, are treated
as multiples of these units.
There have been a number of variants of Kolomogorov's work: {\it Schwarzacher} [1975] described
a process of bed formation that results in a random walk on the integers, {\it Vistelius
and Feigel'son} [1965] allowed different types of sediment to be deposited, {\it Dacey} [1979]
considered both exponential and geometrical probability distributions for the
amount of sediment deposited and eroded, and {\it Strauss and Sadler} [1989] have considered
a continuous version of the random-walk model. These models are generally considered to
be successful at predicting observed bed thickness distributions [{\it Strauss
and Sadler}, 1989].

{\it Tipper} [1983] was
the first to apply the random-walk model to the problem of stratigraphic
completeness: given that deposited sediment is often later eroded,
how much of the depositional history
is preserved in a given stratigraphic section?
{\it Sadler} [1981] obtained a solution
to this problem by
investigating the dependence of sedimentation rate on the time span over which the
sedimentation rate was measured. If the dependence of the sedimentation rate
on time span can be assessed, then for a single stratigraphic section, the ratio of
the overall accumulation rate to the average rate at time span $T$ is the completeness
[{\it Sadler and Strauss}, 1990].
Sadler quantified
the sedimentation rate, $R$, as a power-law function of time span, $T$,
with exponent -0.65: $R\propto T^{-0.65}$. {\it McShea and Raup} [1986] have critically
reviewed Sadler's approach, indicating possible biases in the data he compiled.
Sadler interpreted the decreasing sedimentation
rate with time as the result of including longer and longer hiatuses of erosion or
non-deposition in the average at longer time intervals. Plotnick [1986] introduced
a fractal model for
the length distribution of stratigraphic hiatuses that is consistent with this
interpretation and predicts a power-law dependence of sedimentation rate on time span.
{\it Tipper} [1983], {\it Strauss and Sadler} [1989], and {\it Sadler and Strauss}
[1990] have addressed the issue of stratigraphic completeness with the random-walk
model of sedimentation. The random-walk model predicts a power-law dependence of
sedimentation rate on time with exponent $-\frac{1}{2}$: $R\propto T^{-\frac{1}{2}}$,
giving quite good agreement with Sadler's data.

The time history of sedimentation at a point based on our model is given
in Figure 3.14 and 3.15.
Figure 3.14 is the complete history
of deposition and erosion at a point in the basin. The time series of
deposition and erosion is represented by a fractional Brownian walk with
power spectrum $S(f)\propto f^{-\frac{3}{2}}$. This fractional
Brownian walk represents the elevation
of total height of sediment deposited locally in a fluvial sedimentary
basin, superimposed on a constant rate of subsidence.
The time series is scale-invariant in terms of the nondimensional
sedimentary thickness, $h\sigma /D$, and time, $t\sigma ^{2}/D$;
it is characterized by the single parameter $\sigma /\overline{\eta}$.
If $\sigma /\overline{\eta}$ is small the fluctuations in sedimentation
rate are small compared to the subsidence rate; if $\sigma /\overline{\eta}$
is large the fluctuations are large. For the example given in Figure 3.14, 
$\sigma /\overline{\eta}=0.1$.
Figure 3.15 is produced
from Figure 3.14 by removing any deposited sediment that is subsequently
eroded. In the ``staircase'' plot of Figure 3.15,
beds are defined as a time interval of continuous deposition, i.e. a series of consecutive
timesteps with increasing elevations. Hiatuses are defined as periods in which
no sediment is preserved, i.e. a series
of consecutive timesteps with the same elevation.

We will next discuss the relationship between sedimentation rate and time span
with the stratigraphic model of Plotnick (1986) based on a deterministic
fractal distribution of hiatus lengths. The age of
sediments in this model is given as a function of depth in Figure 3.16a.
As illustrated, the vertical segments (beds) are of equal thickness.
The positions of the transitions from beds to hiatuses are given by a
second-order Cantor set.
Eight kilometers of sediments have been deposited in this model sedimentary
basin in a period of 9\ Myr so that the mean rate of deposition is
$R$ (9 Myr) = 8 km / 9 Myr = 0.89 mm/yr over this period. However, there is
a major unconformity at a depth of 4 km. The sediments immediately above
this unconformity have an age of 3 Ma and the sediments immediately below
it have an age of 6 Ma. There are no sediments in the sedimentary pile
with ages between 3 and 6 Ma. In terms
of the the Cantor set this is illustrated in Figure 3.16b. The line
of unit length is divided into three parts and the middle third,
representing the period without deposition, is removed. The two
remaining parts are placed on top of each other as shown.
 
During the first three million years of deposition (the lower half of
the sedimentary section) the mean rates of deposition are $R$ (3 Myr)=
4 km / 3 Myr = 1.33 mm/yr. Thus the rate of deposition increases as
the period considered decreases. This is shown in Figure 3.16c.
 
There is also an unconformity at a depth of two kilometers. The
sediments immediately above this unconformity have an age of 1 Ma
and sediments below have an age of 2 Ma. Similarly there is an
unconformity at a depth of 6 km, the sediments above this unconformity
have an age of 7 Ma and sediments below an age of 8 Ma. There are
no sediments in the pile with ages between 8 and 7 Ma or between 2 and
1 Ma. This is clearly illustrated in Figure 3.16a. In terms of the
Cantor set, Figure 3.16b, the two remaining line segments of length
1/3 are each divided into three parts and the middle thirds are
removed. The four remaining segments of length 1/9 are placed on
top of each other as shown. During the periods 9 to 8, 7 to 6, 3 to 2,
and 1 to 0 Myr the rates of deposition are $R$ (1 Myr) = 2 km / 1 Myr =
2 mm/yr. This rate is also included in Figure 3.16c.

The rate of deposition clearly has a power-law dependence with respect to the length
of the time interval considered. The results illustrated in Figure 3.16
are based on a second-order Cantor set but the construction can
be extended to any order desired and the power-law results given in
Figure 3.11c would be extended to shorter and shorter time intervals.
 
The sedimentation rate has been calculated in this way based on the
sedimentation history
of Figure 3.15. The results are plotted in Figure 3.17 on a logarithmic
scale. The sedimentation rate has a power-law dependence on time span
with exponent $-\frac{3}{4}$: $R\propto T^{-\frac{3}{4}}$. {\it Sadler and
Strauss} [1990] have shown that the random-walk model results in a power-law
relationship with exponent $-\frac{1}{2}$.
Our result is a better fit to the data of
Sadler (1981) who has compiled measurements of fluvial sedimentation
rates from the geological literature for
time scales of minutes to 100 million years.
His data are plotted in Figure 3.18
where they are averaged in bin sizes with an equal spacing on a logarithmic 
scale.
In this plot we have not included the data on time scales
from 10$^{5}$ to 10$^{8}$ years since these time scales include unconformities
resulting from regressive and transgressive events on active margins.
Variations in
sea level are beyond the scope of the model and it would be
inappropriate to compare the model to sedimentation rates on those time scales.
A least-square linear fit to the
log-log plot yields a slope of $-0.76$.
This result is consistent with the model result given in Figure 3.17.

These results can also be obtained from theoretical fractal relations. Fractional
Brownian walks have the property that the standard deviation
of the time series
has a power-law dependence on time with a fractional exponent $Ha$, the Hausdorff measure:
$\sigma\propto T^{Ha}$ as given by MT equation 3.2. The rate of change of the time series for a given time
interval, $T$, is
then the sedimentation rate $R=\sigma/T\propto T^{Ha-1}$. The power-spectral exponent of a time series
and its Hausdorff measure are related by $\beta =2Ha+1$ (MT equation 5.10).
For the
random-walk model, $\beta=2$, $Ha=1/2$, and the sedimentation rate is then
$R\propto T^{-\frac{1}{2}}$. For the stochastic diffusion model, $\beta=3/2$,
$Ha=1/4$, and $R\propto T^{-\frac{3}{4}}$, in agreement with the numerical results.
 
The dependence of sedimentation rate on time span continues up to time scales
of the Wilson cycle. On time scales of 10$^{5}$-10$^{8}$ years, transgressive and
regressive events
give rise to alternating
periods of deposition and erosion as mentioned previously.
{\it Korvin} [1992] found, using the SEDPAK
simulation package, that alternating periods of deposition and erosion
resulting from sea level
change, combined with the diffusive parameterization of sediment
transport of SEDPAK, resulted in a decreasing sedimentation rate with
increasing time span in the same way that channel avulsion and diffusive
sediment transport results in episodic sedimentation rates on smaller time scales.

\subsection{Bed thicknesses}
Working from our preserved thickness history of Figure 3.10b, we will
define a bed as any consecutive sequence of time units with different
thickness. Conversely, a hiatus is any consecutive sequence of time
units with the same thickness. In this section we will present bed thickness
and hiatus length distributions and compare them with observations and with other
models.
 
{\it Plotnick} [1986] presented the model for discontinuous sedimentation based on
a fractal distribution of hiatus lengths from in Figure 3.15.
The cumulative distribution of hiatus lengths, the number of hiatuses
greater than or equal to a length of time, $T$,
produced by our model is plotted in Figure 3.19. In order to obtain an
accurate curve, we generated 100 synthetic preserved thickness histories
and accumulated the hiatus distributions to obtain Figure 3.19. The distribution
is not fractal. This was at first surprising since a fractal distribution of
hiatuses was used to illustrate how a power-law dependence of sedimentation rate
on time span can occur. However, in the model of Figure 3.16
each bed had the same thickness.
In contrast, as we will show,
the stochastic diffusion model of sedimentation results in bed thicknesses
with an exponential distribution.
Therefore, our observation of a scale-invariant sedimentation rate without
a scale-invariant distribution of hiatuses is not inconsistent with the model
of Figure 3.16 since they result in different bed thickness distributions.
 
The cumulative distribution of bed thicknesses generated by our model is
plotted in Figure 3.20 for the four different values of $\sigma /\overline{\eta}$
indicated next to each distribution.
For synthetic depositional histories with a relatively large $\sigma /\overline{\eta}$,
such as 0.1, no deposition occurs during
most of the history. The result is a small number of beds with a very
skewed distribution. For smaller ratios, more thick beds
appear in the record. The straight line trends of the distributions on a log-linear axis
indicate that the cumulative bed thickness distributions are exponential. The non-cumulative
distribution is also exponential since the cumulative distribution is the integral
of the non-cumulative distribution. Exponential
bed thickness distributions are common in stochastic models of sedimentation [{\it Dacey}, 1979].
Despite reported conclusions
that stochastic models of sedimentation, including those that generate
exponential bed thickness distributions, accurately predict observed bed-thickness distributions
[{\it Mizutani and Hattori}, 1972], we are not aware of any model which predicts the commonly
observed log-normal distribution. This
may be a fundamental weakness of the bed formation
models that
have been proposed to date. Another possibility has been suggested by
{\it Drummond and Wilkinson} [1996]. They have argued that the
observation of log-normal distributions is an artifact resulting from unrecognized
or unrecorded small strata. They propose that exponential distributions are
consistent with the data if the data for the frequencies of the smallest strata
are
considered incomplete and not considered in the distribution fitting.
This is consistent with the conclusion of {\it Muto} [1995] who has presented
the cumulative frequency-thickness distribution of four large turbidite datasets from Japan. He found that an exponential distribution best fit the data.
However, power-law distributions have also been persuasively argued for
the distribution of turbidite beds [{\it Rothman et al.}, 1993].
 
In Figure 3.20, synthetic sedimentation histories with larger values of
sedimentation rate, $\sigma /\overline{\eta}$,
have a more skewed distribution or a steeper slope on
a log-linear scale. This is consistent with the dependence of skew on
sedimentation rate
observed in deep-sea sequences in Italy by {\it Claps and Masetti} [1994]. These
authors published bed-thickness data from three formations in Italy: Ra Stua, Castagne,
and Cismon Valley. The sedimentation
rates for a 1 Ma time scale have been estimated to be 2.5, 1.7, and 0.6 cm/kyr, respectively, for
these sections.
In Figure 3.20 we found that basins which filled
slowly had bed-thickness distributions that were
more skewed than those in basins which filled more quickly.
The cumulative bed-thickness distributions for these sections based on
data that were digitally scanned from {\it Claps and Masetti} [1994] are
presented in Figure 3.21. The model prediction that the skew of the
bed thickness distributions increases from the (a) Ra Stua section to the (b) Castagne and (c)
 Cismon
Valley sections is consistent with the data.

\section{Variability of the earth's magnetic field}
\subsection{Variations of the dipole moment}
\setcounter{equation}{0}
As our third and final example we consider the time series of the earth's 
magnetic field. Paleomagnetic studies show clearly that the polarity
of the magnetic field has been subject to reversals.
{\it Kono} [1971]
has compiled paleointensity measurements of the magnetic field from volcanic
lavas for 0-10 Ma.
He concluded that the distribution of
paleointensity is well approximated by a symmetric binormal
distribution
with mean 8.9x10$^{22}$Am$^{2}$ and standard deviation 3.4x10$^{22}$Am$^{2}$.
A normal distribution is applicable to the field when it is in
its normal polarity and the other when it is in its reversed polarity.

We have utilized three data sets for computing the power spectrum of
the dipole moment of the earth's magnetic field.
They are archeomagnetic data from time scales
of 100 yr to 8 kyr from {\it Kovacheva} [1980], marine sediment
data from the Somali basin from time scales of 1 kyr to 140 kyr from
{\it Meynadier et al.} [1992], and marine sediment data from the Pacific and
Indian Oceans from 20 kyr to 4 Myr from {\it Meynadier et al.} [1994].
The data were published in table form in
{\it Kovacheva} [1980] and obtained from L. Meynadier (personal communication, 1995)
for the marine sediment
data in {\it Meynadier et al.} [1992] and {\it Meynadier et al.} [1994].
Marine sediment data are accurate measures of
relative paleointensity but give no information on absolute intensity.
In order to calibrate marine sediment data, the data must be compared to absolute
paleointensity measurements from volcanic lavas sampled from the same time
period as the sediment record. {\it Meynadier et al.} [1994] has done this
for the composite Pacific and Indian Ocean data set. They have calibrated the
mean paleointensity in terms of the virtual axial dipole moment
for 0-4 Ma as 9x10$^{22}$Am$^{2}$
[{\it Valet and Meynadier}, 1993]. This value is
consistent with that obtained by {\it Kono} [1971] for the longer time
interval up to 10 Ma. Using this
calibration, we calibrated the Somali data with the time interval 0-140 ka from the
composite Pacific and Indian Ocean dataset.
The data from {\it Meynadier et al.} [1994] are plotted in Figure 4.1 as a function of
age in Ma. The last reversal at approximately 730 ka is clearly shown.
We computed the power spectrum of each of the time series with the Lomb
periodogram [{\it Press et al.},
1992].
The compiled spectra are given in Figure 4.2.
The composite sediment record from the Pacific and
Indian Oceans are plotted up to the frequency  
1/(25 kyr). Above this time scale good synchroneity is observed
in the Pacific and Indian Ocean data sets [{\it Meynadier et al.}, 1994]. This
suggests
that non-geomagnetic effects such as variable sedimentation rate
are not significant in these cores above this time scale.
From frequencies of 1/(25 kyr) to 1/(1.6 kyr)
we plot the power spectrum of the Somali data.
From time scales of 1.6 kyr to
the highest frequency we plot the power spectrum of the
data of {\it Kovacheva} [1980].
A least-squares linear regression to the data yields a
slope of $-1.09$ over $4.5$ orders of magnitude. This indicates that
the power spectrum is well approximated as $1/f$ on these time scales.

The power spectrum of secular geomagnetic intensity variations has been
determined to have a $1/f^{2}$ power spectrum between time scales of one
and 100 years [{\it Currie}, 1968; {\it Barton}, 1982;
{\it Courtillot and Le Mouel}, 1988]. This is consistent with the
analysis of {\it McLeod} [1992] who found that the first difference of the annual
means of geomagnetic field intensity is a white noise since
the first differences of a random process with power spectrum $1/f^{2}$
is a white noise.
Our observation of $1/f$ power-spectral behavior above time scales
of approximately 100 yr together with the results of {\it Currie} [1968] and
{\it Barton} [1982]
suggests that there is a crossover from $1/f$ to $1/f^{2}$ spectral
behavior at a time scale of approximately
one hundred years.

\subsection{Reversal record} 
We will now
show that the statistics of the reversal record are
consistent those with of a binormal $1/f$ noise paleointensity record which
reverses each time the intensity crosses the zero value.
We will compare the polarity length distribution
and the clustering of reversals between synthetic reversals produced with
$1/f$ noise intensity variations and the
reversal history according to {\it Harland et al.} [1990] and {\it Cande and Kent}
[1992a,1995].

First we consider the polarity length distribution of the real reversal
history. The polarity length distribution calculated from the chronology of
{\it Harland et al.} [1990] is given as the solid line in Figure 4.3. The polarity
length distribution is
the number of interval lengths
longer than the length plotted on the horizontal axis.
A reassessment of the magnetic anomaly data
has been performed by {\it Cande and Kent} [1992a,1995] to obtain an alternative
magnetic
time scale. The polarity length distribution of their time scale normalized to the same length as the {\it Harland
et al.} [1990] time scale, is presented as the dashed curve. The two distributions
are nearly identical.
These plots suggest that the polarity length distribution is better fit
by a power law for large polarity lengths than by an exponential distribution,
as first suggested by {\it Cox} [1968].
The same conclusion has been reached by {\it Gaffin} [1989] and {\it Seki and Ito}
[1993].
 
The third curve, plotted with a dashed-dotted line, represents
the polarity length distribution estimated from the magnetic time scale between
C1 and C13 with ``cryptochrons'' included and scaled to the length of the
{\it Harland et al.} [1990] time scale. Cryptochrons are small variations
recorded in the magnetic anomaly data that may either represent variations in
paleomagnetic intensity or short reversals [{\it Blakely}, 1974;
{\it Cande and Kent}, 1992b].
Cryptochons occur with a time scale at
the limit of temporal resolution of the reversal record from
magnetic anomalies of the sea floor.
The form of the polarity length distribution estimated from the record between
C1 and C13
including cryptochrons is not representative of the entire reversal history
because of
the variable reversal rate which
concentrates many short polarity intervals in this time period. However, this
distribution enables us to estimate the temporal resolution of the reversal
record history. The distribution estimated from C1 to C13
has many more short polarity intervals than those of the full reversal
history starting at a reversal length of 0.3 Myr. Above a time
scale of 0.3 Myr the magnetic time scale is nearly complete. Below it many
short polarity intervals may be unrecorded.

To show that this distribution is consistent with binormal
$1/f$ noise intensity variations, we have generated
synthetic Gaussian noises with a power spectrum porportional to $1/f$,
a mean value of 8.9x10$^{22}$Am$^{2}$ and a standard deviation of 3.4x10$^{22}$Am$^{2}$ as
obtained by {\it Kono} [1971], representative of the field intensity in one
polarity state. The synthetic noises were generated using
the Fourier-domain filtering technique discussed in MT section 2.1. 
An example is shown as Figure 4.4.
In order to construct a binormal intensity distribution from the synthetic
normal distribution, we inverted every other polarity interval to the
opposite polarity starting from its minimum value below the zero intensity
axis and extending to its next minimum below the zero. The result of this
procedure on the Gaussian, $1/f$ noise of Figure 4.4 is presented 
in Figure
4.5.
Its irregular polarity lengths are similar to that in the marine sediment
data of Figure 4.1.
 
The operation of reversing the paleomagnetic intensity when it crosses the
zero intensity value is consistent with models of the geodynamo as a
system with two symmetric attracting states of positive and negative
polarity such as the Rikitake disk dynamo. Between reversals, the
geomagnetic field fluctuates until a fluctuation large enough occurs
to cross the energy barrier into the other basin of attraction. {\it Kono} [1987]
has explored the statistical similarity between the Rikitake disk dynamo
and the distribution of paleointensity. Our
construction of the binormal $1/f$ noise is consistent with this model.
 
We have
computed the distributions of lengths between successive reversals
for twenty synthetic noises scaled to length 169 Ma, the length of the reversal
chronology, and averaged the
results in terms of the number of reversals. The results are plotted
as the solid curve along with the {\it Harland et al.} [1990]
time scale (dashed curve) in Figure 4.6. The dots in Figure 4.6 
are the maximum and minimum values
obtained in the twenty synthetic reversal
chronologies for each reversal rank, thus
representing 95\% confidence intervals. The shape
of the synthetic polarity length distribution is very similar to the {\it Harland
et al.} [1990] time scale.
The synthetic polarity length distribution matches the {\it Harland et al.}
[1990] time scale 
within the 95\% confidence interval
over all time scales plotted except for the Cretaceous
superchron, which lies slighly outside of the 95\% confidence interval,
and reversals
separated by less than about 0.3 Myr.
The overprediction of very short
reversals could be a limitation of the model or a result of the incompleteness
of the reversal record for short polarity intervals.
As mentioned, the temporal resolution of the magnetic time scale inferred from
magnetic anomalies is approximately 0.3 Myr. We conclude that the
polarity length distribution produced from binormal $1/f$ intensity variations
are consistent with the observed
polarity length distribution for all time scales for which the reversal record
is complete.

We next consider whether the agreement illustrated
in Figure 4.6 is unique to $1/f$ noise. We have computed polarity
length distributions using binormal intensity variations
with power spectra $f^{-0.8}$ and $f^{-1.2}$. These results along with
the $1/f$ result from Figure 4.6 are given in Figure 4.7.
The shape of the polarity length distribution is very sensitive to the
exponent of the power spectrum. A slight increase in the magnitude of
the exponent results in many more long polarity intervals than with
$1/f$ noise. We conclude that the agreement in Figure 4.6 between the
synthetic reversal distribution and the true reversal history is unique
to $1/f$ noise and provides strong evidence that the dipole moment has
$1/f$ behavior up to time scales of 170 Myr.
 
A binormal, $1/f$ noise geomagnetic field variation is consistent with the
qualitative results of {\it Pal and Roberts} [1988] who found an
anticorrelation
between reversal frequency and paleointensity. This anticorrelation is
evident in the synthetic $1/f$ noise of Figure 4.5. During the time intervals
of greatest average paleointensity the reversal rate is lowest.
 
In addition to the broad distribution of polarity lengths, the reversal
history is also characterized by a clustering of reversals. This behavior
has been quantified with the reversal rate. The reversal rate has
been relatively high
from 0-20 Ma and has decreased gradually going back in history
to the Cretaceous superchron.
An alternative approach to quantifying the clustering of reversals is
with the pair-correlation function.
The pair-correlation function $C(t)$ is the
number of pairs of reversals whose separation is between $t$ and $t+\Delta t$,
per unit time [{\it Vicsek}, 1992].
The pair-correlation function for a
set of points can be compared to that for a Poisson process to detect
non-random clustering.
The pair-correlation function analysis is more
appropriate for comparison of the reversal history to the synthetic reversal
history generated by a stochastic model 
since a stochastic model cannot predict behavior in time, such as
when the reversal rate is large or small.
However, a stochastic model
may accurately reflect the extent to which small polarity intervals
are followed by small polarity intervals and long intervals by long intervals
as quantified with the pair-correlation function.
 
The pair-correlation function of
reversals according to the {\it Harland et al.} [1990] and {\it Cande and Kent} [1992a,1995]
reversal history
are shown in
Figure 4.8 as filled and unfilled circles, respectively.
Also presented in
Figure 4.8 is the pair-correlation function for a synthetic
reversal data set based on binormal $1/f$ noise dipole moment variations
(boxes) and for a Poisson process (triangles).
The functions are offset so that they may be placed on the same graph.
The Poisson process was constructed with 293 points, the same number of reversals
as the {\it Harland et al.} [1990] time scale,
positioned with uniform probability on the interval between
0 and 170 Ma. The Poisson process yields a correlation function
independent of $t$.
The real and synthetic reversal histories
variations exhibit significant clustering with more pairs of points at
small separation and fewer at large separations than for a Poisson process.
Straight-line fits of the form $C(t)\propto t^{-\alpha}$ were obtained.
The purpose of this was to show that
similar clustering is observed in the real and synthetic reversals. The exponents
of the {\it Harland et al.} [1990], {\it Cande and Kent} [1992a,1995],
and synthetic reversals are -0.39, -0.31, and -0.42, respectively,
indicating close agreement between the model and real reversals.

\subsection{Inclination and declination data}
Power spectral analyses of inclination and declination data have also been
carried out. We obtained time series data for inclination and declination
from lake sediment cores in the Global Paleomagnetic Database [{\it Lock
and McElhinney}, 1992]. The core with the greatest number of data points
was from Lac du Bouchet [{\it Thouveny et al.}, 1990]. The inclination data
from this data set is plotted in Figure 4.9. The power spectrum
of the inclination and declination at Lac du Bouchet
estimated with the Lomb Periodogram is
presented in Figure 4.10. We associate the spectra with a flat spectrum
below a frequency of $f\approx$ 1/(3 kyr) and a constant spectrum
above a frequency of $f\approx$
1/(500 yr). From frequencies of $f\approx$ 1/(3 kyr) to
$f\approx$ 1/(500 yr) the inclination and declination are Brownian walks
with $S(f)\propto f^{-2}$. Spectral analyses of inclination data from
five other sediment cores were calculated. These spectra are presented in
Figure 4.11. The spectra correspond, from top to bottom, to cores from
Anderson Pond [{\it Lund and Banerjee}, 1985], Bessette Creek [{\it Turner et al.},
1982], Fish Lake [{\it Verosub et al.}, 1986],
Lake Bullenmerri [{\it Turner and Thompson}, 1981],
and Lake Keilambete [{\it Barton and McElhinny}, 1985]. Since the data sets have
fewer points there is more uncertainty in the spectra and they are characterized
by greater variability between adjacent frequencies. The spectra have the
same form, within the uncertainty of the spectra, as that associated with the
spectra from Lac du Bouchet. These results suggest that 3 kyr and 500 yr are
characteristic time scales of geodynamo behavior.
Variations in inclination and declination are associated with changes in
the non-dipole components of the field.
Therefore, the autocorrelation or decay time of the quadrupole moment is the
maximum time scale for correlated fluctuations of inclination and declination
to occur. The autocorrelation time of the quadrupole moment has been estimated
by {\it McLeod} [1996] to be 1.6 kyr. This is within a factor of two
of the 3 kyr time scale above which variations in inclination and declination
are observed to be uncorrelated in the spectra of Figures 4.10 and 4.11.

Many analyses of variations
in paleointensity of the earth's magnetic field concentrate on identifying
characteristic time scales of variation. Many such characteristic
time scales have been identified. {\it Valet and Meynadier} [1993] suggested,
based on the same sediment core data analyzed in this paper, that the
earth's magnetic field regenerates following a reversal on a time scale of
a few thousand years and then decays slowly on a time scale of 0.5 Myr
before the next reversal. They termed this an ``asymmetric saw-tooth'' pattern.
More recent data have shown that the ``asymmetric saw-tooth'' is not
a robust pattern.
Longer cores show a slow decay
preceding a reversal to be rare [{\it Tauxe and Hartl}, 1997]. Moreover, {\it Laj et al.}
[1996] has shown that the magnetic field does not always regenerate quickly
after a reversal.
{\it Thibal et al.} [1995] have quantified the rate of decrease in field intensity
preceeding a reversal and found it to be inversely proportional to the length
of the polarity interval. The authors concluded from this
that the length of the reversal
was predetermined. Such behavior is not indicative of a predetermined polarity
length. This can be concluded by considering the null hypothesis that variations
in the field are characterized by any stationary random process.
By definition, a stationary time series has a variance which is independent
of the length of the series.
The average rate of change of the time series over a time interval
will then be a constant value divided by the interval of time, i.e. inversely
proportional to time interval. Therefore, any stationary random function satisfies
the relationship that {\it Thibal et al.} [1995] observed.
 
In the power-spectral analyses
of geomagnetic variations inferred from sediment cores performed by {\it Lund et al.} [1988],
{\it Meynadier et al.}
[1992], {\it Lehman et al.}
[1996], and {\it Tauxe and Hartl} [1997]
dominant periodicities
in the record were identified and proposed as characteristic time scales
of geodynamo behavior.
However, it must be emphasized that any finite length record will
exhibit peaks in its power spectrum even if the underlying
process is random such as a $1/f$ noise. Periodicity tests such
as those developed by {\it Lees and Park} [1995] need to be applied to data
in order to assess the probability that a peak in a spectrum is statistically
significant. The periodicity tests developed by {\it Lees and Park} [1995]
are especially valuable because they do not depend on a particular model of the
stochastic portion of the spectrum.
Some of the periodicity tests that have been used in
the geomagnetism literature assume forms for the stochastic portion of the
spectrum that are not compatible with the $1/f$ process we have identified.
See {\it Mann and Lees} [1996] for an application
of these techniques to climatic time series.
 
It is generally believed that secular geomagnetic variations are the result of
internal dynamics
while longer time scale phenomena such as variations in the reversal
rate are controlled by variations in boundary conditions at
the core-mantle boundary (CMB) [{\it McFadden and Merrill}, 1995].
However, our
observation of continuous $1/f$ spectral behavior from time scales of 100 yr
to 170 Myr suggests that a single process controls variations in geomagnetic
intensity over this range of time scales. In Section 4.4
we consider a
model for geodynamo behavior which reproduces the $1/f$ dipole moment
variations over a wide range
of time scales and
exhibits many of the other features of geomagnetic variability
we have identified.

\subsection{Model for geomagnetic variations}
There has been great interest in $1/f$ noise processes in the physics
literature for many years [{\it Weissman}, 1988].
One model of $1/f$ noise is a
stochastic
process comprised of a superposition of modes with exponential decay characterized
by different
time constants. The time constant for a stochastic
process is defined through its autocorrelation function $a(\tau)$.
For a stochastic process with a single time constant $\tau_{o}$
the autocorrelation function is given by $a(\tau)=e^{-\frac{\tau}{\tau_{o}}}$.
The power spectrum of such a process is, by the Weiner-Khinchine theorem, the
Fourier transform of the autocorrelation function:
\begin{equation}
S(f)\propto\frac{\tau_{o}}{1+(2\pi f)^{2}}
\label{lorentzian}
\end{equation}
This is a Lorentzian spectrum with Brownian walk behavior ($S(f)\propto f^{-2}$)
for time scales small
compared to $\tau_{o}$ and white noise
behavior ($S(f)=$ constant) above the characteristic time constant.
If the stochastic process is composed of a superposition of modes with
time constants following a distribution
$D(\tau_{o})\propto\tau_{o}^{-1}$, where the $D(\tau_{o})\Delta\tau_{o}$ is the
net variance contributed by modes between $\tau_{o}$ and $\tau_{o}+\Delta\tau_{o}$,
then a $1/f$ spectrum results over a range of frequencies
[{\it van der Ziel}, 1950; {\it Weissman}, 1988].
Such a distribution of exponential
time constants has
been documented for the earth's magnetic field by {\it McLeod} [1996].

{\it McLeod} [1996]
calculated the autocorrelation of each degree of the geomagnetic field
during the last eighty years. The autocorrelation functions that he computed had
an exponential dependence on time with degree-dependent time constants
$\tau_{o}\propto n^{-2}$.
This behavior
is consistent with a diffusion process.
{\it McLeod} [1996]
attributed this autocorrelation structure
to a simple model of
the geomagnetic field in which the field was stochastically
generated with a balance between field regeneration and diffusive decay 
across a magnetic boundary layer.
One way to model such a stochastic diffusion process
is with a two-dimensional diffusion equation driven by random noise:
\begin{equation}
\frac{\partial B_{z}}{\partial t}=D\nabla^{2} B_{z} +\eta(x,y,t)
\end{equation}
where $B_{z}$ is the axial component of the magnetic field at a point inside the
core and
$\eta(x,y,t)$ is a Gaussian white noise representing random amplification
and destruction of the field locally by dynamo action.
To this equation we add a term equal to $c(p-B_{z,tot})$:
\begin{equation}
\frac{\partial B_{z}}{\partial t}=D\nabla^{2} B_{z} +\eta(x,y,t)+c(p-B_{z,tot})
\label{uh}
\end{equation}
where $c$ is a constant, $B_{z,tot}$
is the dipole moment integrated over all space, and
$p$ is +1 if the dipole moment of the field outside the core-mantle boundary
is positive and -1 if the dipole moment outside the core-mantle boundary is
negative.
The effect of this term is
to create two basins of attraction (polarity states) within which the dipole
field fluctuates around an intensity of +1 or -1
until a fluctuation large enough occurs to cross the barrier
to the other basin of attraction. This term
could be the result of a conservation of magnetic energy for the combination
of the poloidal and toroidal fields
such that when the poloidal dipole field intensity
is low the toroidal field
intensity, which is unobservable outside the core and not explicitly modeled in
equation \ref{uh},
is high and dynamo action is intensified, repelling the poloidal
field away from a state of low dipole intensity.
 
In our model the core is
modeled as a two-dimensional circular region of
uniform diffusivity (the fluid outer core)
surrounded by
an infinite region with small but finite
diffusivity and the boundary condition that $B_{z}$ approach zero as
$r$ approaches zero where $r$ is 
the radial distance from the center of the earth.
The diameter of the inner circular region is the diameter of the
core-mantle boundary.
 
This model has been simulated by computer using finite differencing of
the model equation on a two-dimensional lattice. It has been 
studied in terms of the distribution of values and power spectrum
of the dipole moment
and the
power spectrum of the angular deviation from the dipole field.
The dipole field from the simulation is
plotted in Figure 4.12. The field clearly undergoes reversals with a broad
distribution of polarity interval lengths. Figure 4.13 represents the
dipole distribution of 10 simulations (solid curve) along with the fit to a binormal
distribution (dashed curve). A binormal distribution fits the data well.
The slight asymmetry is the result of this particular model run spending
slightly more time in the negative polarity state than in the positive polarity
state. Model outputs were generated which showed asymmetry in the other direction.

The average power spectrum of time series of the dipole field from 25
simulations is presented in Figure 4.14. The spectrum has a low-frequency
spectrum $S(f)\propto f^{-1}$ and a high-frequency spectrum $S(f)\propto f^{-2}$.
This is identical to the spectrum observed in sediment cores and historical
data discussed earlier in the chapter. The crossover time scale is the diffusion time
across the diameter of the core, estimated to be between
10$^{3}$ yr [{\it Harrison and Huang}, 1990] and 10$^{4}$ yr [{\it McLeod}, 1996].
These values are somewhat higher than the time scale of 10$^{2}$
yr identified as the crossover in the sediment core and historical data.

Figure 4.15 shows the average power spectrum of the angular displacement from
the dipole from 25 simulations. The spectrum has a
high-frequency region $S(f)\propto f^{-2}$ which slowly flattens out to
a flat spectrum at low frequencies. This is nearly consistent with the
spectra of inclination and declination from lake sediment time series shown
in Figures 4.10 and 4.11.
The measured value of the crossover from white noise to Brownian walk behavior
in the lake sediment power spectra is 3 kyr. This value is consistent
with estimates of 10$^{3}$ to 10$^{4}$ years for the diffusion time across the core
from {\it Harrison and Huang} [1990] and {\it McLeod} [1996]. A major discrepancy between
the model and the observed spectrum is the absence of a flattening out of the
spectrum of angular displacement at high frequencies in the model calculation.

\section{Other Applications}
Self-affine time series occur in many other areas of Earth science. 
For example, topographic profiles are Brownian walks [{\it Turcotte}, 1987].   
{\it Pelletier} [1997c] has shown that a model of topography governed by 
the diffusion equation with the diffusivity a function of discharge predicts
both the Brownian walk variations and the log-normal distribution of topography. 
Branching river networks with statistics identical to those of real river
networks were also obtained. Gravity fields also exhibit power-law power spectra
[{\it Turcotte}, 1987; {\it Passier and Sneider}, 1995]. These power spectra
have been interpreted as resulting from random density anomalies in the mantle
[{\it Lambeck}, 1976].
A related problem to the fractal structure of topography
which may also exhibit self-affinity is sediment loads
in rivers. 
{\it Plotnick and Prestegaard} [1993] have obtained time series data
for sediment loads in rivers on time scales of minutes to days. They applied
both the rescaled-range technique and power-spectral analysis to show that the time
series are approximately self-affine. 

{\it Tjemkes and Visser} [1987] have performed power-spectral analyses
on the horizontal variability of temperature, humidity, and cloud water
in the atmosphere. They found that different power-law behaviors were 
applicable over well-defined wave number ranges. These results are important
for understanding the variability of the atmosphere and for
improved characterization of these fields for inputs into large scale models
of the climate system [{\it IPCC}, 1995].  
The TOPEX/POSEIDON project has provided data on sea surface height
with global coverage with a 10-day sampling interval. {\it Wunsch and Stammer} [1995]
have shown that sea surface height has self-affine behavior in both space and 
time with three different values of $\beta$ characterizing the variability
over different wave number ranges. Variability in sea surface height has been
modeled using the potential vorticity equation with stochastic forcing to
represent variable wind conditions [{\it Muller}, 1996]. 
These techniques are very similar to
the stochastic partial differential equations discussed in Sections 2-4.
{\it Hsui et al.} [1993] have shown that sea level variations are a self-affine
time series on time scales of 10$^{4}$ to 10$^{8}$ yr. 
Since sea level variations determine the major unconformities of the 
stratigraphic record, the record of the earth's history is 
determined by self-affine behavior. 
  
Self-affine time series have applications in other fields.
It has long been recognized that spatial variations in plankton abundance in the
oceans are self-affine. This has been determined by performing power-spectral
analysis on remotely-sensed data for plankton along one-dimensional
transects [{\it Platt and
Denman}, 1975]. Plankton variability has been modeled using stochastic 
diffusion equations similar to those presented in this paper [{\it Fasham}, 1978]. 
Diffusion is used to model ocean mixing and stochastic
terms are introduced to model the effects of local environmental variations that
affect the population growth rate
such as variations in light intensity and nutrient concentration.
Power spectral analyses have also
been performed on vegetation densities [{\it Palmer}, 1988]. The time series
were observed to have power-law power spectra. 
{\it Sugihara and May} [1990] and {\it McKinney and Frederick} [1992] 
have applied the self-affinity of population abundance in time to 
assessing the probability of extinction. They argued that populations with
stronger correlations in variability, characterized by larger values of
$\beta$ or $Hu$, have greater fluctuations in population size and have a
higher probability of extinction.

Self-affine time series with $\beta\approx 1$ 
are also observed in traffic flows [{\it Musha and
Higuchi}, 1976]. This behavior is reproduced in lattice gas models which
move cars around on a lattice according to simple interaction rules
that prevent cars from occupying the same space and that
are driven by a random input of cars into the lattice [{\it Takayasu
and Takayasu}, 1993]. 
$1/f$ noise has also been observed in the the density of internet traffic.  
This observation
may have important implications for the design and testing of network software
and services. 


\newpage

Alexander, L.L., Geologic evolution and stratigraphic controls on fluid
flow of the Eugene Island Block 330 Mini Basin, offshore Louisiana,
Ph.D. dissertation, Cornell Univ., Ithaca, N. Y., 1995.

Allen, P.A., and J.R. Allen, {\it Basin Analysis: Principles and Applications},
Blackwell Sci., Cambridge, Mass., 1990.

Arking, A., and D. Ziskin, Relationship between clouds and sea surface temperatures
in the western tropical Pacific, {\it J. Clim., 7}, 988, 1994.

Barnett, T.P., A.D. Del Genio, and R.A. Ruedy, Unforced decadal fluctuations
in a coupled model of the atmosphere and ocean mixed layer, {\it J. Geophys. Res., 97},
7341-7354, 1992.

Barton, C.C., and C.H. Scholz, The fractal size and spatial
distribution of hydrocarbon accumulations: implications for resource
assessment and exploration strategy, edited by C.C. Barton and P.R. LaPointe,
{\it Fractals in Petroleum Geology and Earth Sciences}, pp. 13-34, Plenum,
New York, 1995.

Barton, C.E., Spectral analysis of palaeomagnetic time series
and the geomagnetic spectrum, {\it Phil. Trans. R. Soc. London A, 306},
203-209, 1982.

Berg, R.R., Point bar origin of Fall River sandstone reservoirs, northeast
Wyoming, {\it Am. Assoc. Pet. Geol. Bull., 52}, 2116-2122, 1968.

Blakely, R.J., Geomagnetic reversals and crustal spreading rates during the
Miocene, {\it J. Geophys. Res., 79}, 2979-2985, 1974.

Booy, C., and L.M. Lye, A new look at flood risk determination, {\it
Water Resour. Bull., 25}, 933-943, 1989.

Bras, R.L., and I. Rodriguez-Iturbe, {\it Random Functions and Hydrology},
Addison-Wesley, Reading, Ma., 1985.

Cande, S.C., and D.V. Kent, A new geomagnetic polarity time scale
for the Late Cretaceous and Cenozoic, {\it J. Geophys. Res., 97},
13,917-13,951, 1992a.

Cande, S.C., and D.V. Kent, Ultrahigh resolution marine magnetic anomaly
profiles
: a record of continuous paleointensity variations?, {\it J. Geophys. Res.,
97}, 15075-15083, 1992b.

Cande, S.C., and D.V. Kent, Revised calibration of the geomagnetic polarity
timescale for the Late Cretaceous and Cenozoic, {\it J. Geophys. Res., 100},
6093-6095, 1995.

Claps, M., and D. Masetti, Milankovitch periodicities recorded in Cretaceous
sequences from the Southern Alps (Northern Italy), in {\it Orbital Forcing
and Cyclic Sequences}, edited by P.L. DeBoer and D.L. Smith, pp. 99-107,
Blackwell Sci.,
Cambridge, Mass., 1994.

Coleman, J.M., and D.B. Prior, Deltaic environments of deposition, in
{\it Sandstone Depositional Environments},
edited by P.A. Scholle and D. Spearing, {\it AAPG Mem., 31}, 139-178, 1982.

Courtillot, V., and J.L. Le Mouel, Time variations of the earth's magnetic
field: From daily to secular, {\it Ann. Rev. Earth Plan. Sci., 16},
389-476, 1988.

Cox, A., Lengths of geomagnetic polarity intervals, {\it J. Geophys. Res.,
73}, 3247-3260, 1968.

Currie, R.G., Geomagnetic spectrum of internal origin and lower
mantle conductivity, {\it J. Geophys. Res., 73}, 2779-2768, 1968.

Curry, W.H., and W.H. Curry III, South Glennock oil field, Wyoming: A
pre-discovery thinking and post-discovery description, in
{\it Stratigraphic Oil and Gas Fields}, edited by R.E. King, {\it Mem. Am.
Assoc. Pet. Geol., 15}, 415-427, 1972.

Dacey, M.F., Models of bed formation, {\it Math. Geol., 11}, 655-668, 1979.

Deshpande, A., P.B. Flemings, and J. Huang, Quantifying lateral heterogeneities
in fluvio-deltaic sediments using 3-D reflection seismic data: Offshore Gulf
of Mexico, {\it J. Geophys. Res.}, in press, 1997.

Dolan, S.S., C.J. Bean, and B. Riollet, The broad-band fractal nature 
of heterogeneity in the upper crust from petrophysical logs, submitted, 1996.

Drummond, C.N., and B.H. Wilkinson, Stratal thickness frequencies
and the
prevalence of
orderedness in stratigraphic sequences, {\it J. Geol., 104}, 1-18, 1996.

Dunne, T., K.X. Whipple, and B.F. Aubry, Microtopography of hillslopes and
initiation of channels by Horton overland flow, in {\it Natural and AnthropogenicInfluences in Fluvial Geomorphology: The Wolman Volume, Geophys. Monogr. Ser.},
vol. 89, edited by J.E. Costa et al., pp. 27-44, AGU, Washington, D. C., 1995.

Family, F., Scaling of rough surfaces: Effects of surface diffusion,
{\it J. Phys. A Math. Gen., 19}, L441-L446, 1986.

Fasham, M. J. R., The statistical and mathematical analysis of plankton patchiness,
{\it Oceanogr. Mar. Biol. Ann. Rev., 16}, 43-79, 1978.

Gaffin, S., 1989. Analysis of scaling in the geomagnetic polarity reversal record,
{\it Phys. Earth Plan. Inter.}, {\bf 57}, 284-290.

Garcia, R., Depositional systems and their relation to gas accumulation in
Sacremento Valley, California, {\it AAPG Bull., 65}, 653-674,
1981.

Garrett, A.E., Vertical eddy diffusivity in the ocean interior, {\it J.
Mar. Res., 42}, 359-393, 1984.

Ghil, M., Theoretical
climate dynamics: An introduction, in {\it Turbulence
and Predicability in Geophysical Fluid Dynamics and Climate Dynamics}, edited by
M. Ghil, North Holland, Amsterdam, 1983.

Gomes da Silva, L.M., and D.L. Turcotte,
A comparison between Hurst and
Hausdorff measures derived from fractional time series, {\it Chaos, Solitons
and Fractals, 4}, 2181-2192, 1994.

Harland, W.B., Cox, A., Llewellyn, P.G., Pickton, C.A.G., Smith, A.G. \&
Walters, R., {\it A Geologic Time Scale 1989},
Cambridge University Press,
London, 1990.

Harrison, C.G.A., and Q. Huang, Q., Rates of change of the Earth's magnetic
field measured by recent analyses, {\it J. Geomag. Geoelectr., 42},
897-928, 1990.

Harvey, L.D.D., and S.H. Schneider, Transient climate response to
external forcing on $10^{0}$-$10^{4}$ year time scales Part I:
Experiments with globally averaged, coupled, atmosphere and ocean energy
balance models, {\it J. Geophys. Res., 90}, 2191-2205, 1985.

Hays, J.D., J. Imbrie, and N.J. Shackleton, Variations in the earth's orbit:
Pacemaker of the ice ages, {\it Science, 194}, 1121-1132, 1976.

Hewett, T.A., Fractal distribution of reservoir heterogeneity
and their influence of fluid transport, {\it SPE Prof. Pap.} 15386,
Soc. of Pet. Eng., Richardson, Tex., 1986.

Hoffert, M.I., A.J. Callegari, and C.-T. Hsieh, The role of deep sea
heat storage in the secular response to climatic forcing, {\it J. Geophys.
Res., 85}, 6667-6679, 1980.

Hofmann, D.J., and J.M. Rosen, On the prolonged lifetime of the
El Chichon sulfuric acid aerosol cloud, {\it J. Geophys. Res., 92}, 9825-9830, 1987.

Holliger, K., Upper-crustal seismic velocity heterogeneity as derived from a variety
of P-wave sonic logs, {\it Geophys. J. Int., 125}, 813-829, 1996.

Hsui, A.T., K.A. Rust, and G.D. Klein, A fractal analysis of Quaternary,
Cenozoic-Mesozoic, and late Pennsylvanian sea-level changes, {\it J. Geophys. Res.,
98}, 21963-21967, 1993.

Intergovernmental Panel on Climate Change, {\it Climate Change, The IPCC
Scientific Assessment}, edited by J.T. Houghton and B.A. Callendar, Cambridge Univ.
Press, New York, 1995.

Janosi, I.M., and G. Vattay,
Soft turbulent state of the atmospheric boundary
layer, {\it Phys. Rev. A, 46}, 6386-6389, 1992.

Jouzel, J., C. Lorius, J.R. Petit, C. Genthon, N.I. Barkov, V.M. Kotlyakov,
Vostok ice-core: A continuous isotope temperature record over
the last climatic cycle (160,000 years), {\it Nature, 329}, 403-407, 1987.

Jouzel, J., and D. Merlivat, Deuterium and oxygen 18 in precipiation:
Modeling of the isotopic effects during snow formation, {\it J. Geophys.
Res., 89}, 11,749-11,757, 1984.

Kagan, Y.Y., and L. Knopoff, Spatial distribution of earthquakes: The
two-point correlation function, {\it Geophys. J. R. Astron. Soc., 62}, 303-320,
1980.

Kendall, D.R., and J.A. Dracup, A comparison of index-sequential and
AR(1) generated hydrologic sequences, {\it J. Hyd., 122}, 335-352, 1991.

Kerr, R.A., Climate control: How large a role for orbital variations?,
{\it Science, 201}, 144-146, 1978.

Kolmogorov, A.N., Solution of a problem in probability theory
connected with the problem of the mechanism of stratification, {\ Trans. Am.
Math. Soc., 53}, 171-177, 1951.

Kondev, J., and C.L. Henley, Geometrical exponents of contour loops on random
gaussian
surfaces, {\it Phys. Rev. Lett., 74}, 4580-4583, 1995.

Kono, M., Intensity of the earth's magnetic field during the Pliocene and Pleistocene
in relation to the amplitude of mid-ocean ridge magnetic anomalies, {\it Earth and Plan.
Sci. Lett., 11}, 10-17, 1971.

Kono, M., Rikitake two-disk dynamo and paleomagnetism, {\it Geophys. Res. Lett.,
14},
21-24, 1987.

Korvin, G., {\it Fractal Models in the Earth Sciences}, Elsevier, Amsterdam, 1992.

Kovacheva, M., Summarized results of the archeomagnetic investigation of the
geomagnetic field variation for the last 8000 yr in south-eastern Europe,
{\it Geophys. J. R. Astr. Soc., 61}, 57-64, 1980.

Laj, C., C. Kissel, and I. Lefevre, Relative geomagnetic field intensity 
and reversals from Upper Miocene sections in Crete, {\it Earth Plan. Sci.
Lett., 141}. 67-78, 1996.  

Lambeck, K., Lateral density anomalies in the upper mantle, {\it J. Geophys.
Res., 81}, 6333-6340, 1976.

Landwehr, J.M., and N.C. Matalas, On the nature of persistence in
dendrochronologic records with implications for hydrology, {\it J. Hyd.,
86}, 239-277, 1986.

Lees, J.M., and J. Park, Multiple-taper spectral analysis: A stand-alone
C-subroutine, {\it Computers \& Geology}, 21, 199-236, 1995. 

Lehman, B., C. Laj, C. Kissel, A. Mazaud, M. Paterne, and L. Labeyrie, Relative
changes of the geomagnetic field intensity during the last 280 kyear from piston
cores in the Acores area, {\it Phys. Earth Plan. Int., 93}, 269-284, 1996.

Lock, J., and M.W. McElhinney, The Global Paleomagnetic Database: design,
installation, and use with ORACLE, {\it Surveys in Geophysics, 12},
317-506, 1992.

Lund, S.P., and S.K. Banerjee, The paleomagnetic record of Late 
Quaternary secular variation from Anderson Pond, Tennessee, {\it Earth
Plan. Sci. Lett., 72}, 219-237, 1985.

Lund, S.P., J.C. Liddicoat, K.R. Lajoie, T.L. Henyey, and S.W. Robinson,
Paleomagnetic evidence for long-term (10$^{4}$ year) memory
and periodic behavior in the earth's core dynamo process, {\it Geophys.
Res. Lett., 15}, 1101-1104, 1988.    

Manabe, S., and R.J. Stouffer, Low-frequency variability of surface
air temperature in a 1000-year integration of a coupled
atmosphere-ocean-land surface model, {\it J. Climate, 9}, 376-393, 1996.

Manley, G., Central England temperatures: monthly means 1659-1973, 
{\it Quat. J. Roy. Met. Soc., 100}, 389-405, 1974.  

Mann, M.E., and J.M. Lees, Robust estimation of background noise and signal
detection in climatic time series, {\it Climatic Change, 33}, 409-445, 1996.

Matteucci, G., Analysis of the probability distribution of the
late Pleistocene climatic record: Implications for model validation, {\it Clim. Dyn.,
5}, 35-52, 1990.

McFadden, P.L., and R.T. Merrill, History of the Earth's magnetic field
and possible connections to core-mantle boundary processes, {\it J. Geophys.
Res., 100}, 307-316, 1995. 

McKinney, M.L., and D. Frederick, Extinction and population dynamics:
New methods 
and evidence from Paleogene foraminifera, {\it Geology, 20}, 343-346, 1992.

McLeod, M.G., Signals and noise in magnetic observatory annual means:
Mantle conductivity and jerks, {\it J. Geophys. Res., 97}, 17,261-17,290, 1992. 

McLeod, M.G., Spatial and temporal power spectra of the geomagnetic field,
{\it J. Geophys. Res., 101}, 2745-2763, 1996.

Meynadier, L., J.-P. Valet, F.C. Bassonot, N.J. Shackleton, and
Y. Guyodo, Asymmetrical saw-tooth pattern of the geomagnetic field
intensity from equatorial sediments in the Pacific and Indian oceans,
{\it Earth Plan. Sci. Lett., 126}, 109-127, 1994.

Meynadier, L., J.-P. Valet, R. Weeks, N.J. Shackleton, and V.L. Hagee,
Relative geomagnetic intensity of the field during the last 140 ka, {\it
Earth Plan. Sci. Lett., 114}, 39-57, 1992.

Mitchell Jr., J.M., An overview of climatic variability and its causal 
mechanisms, {\it Quat. Res., 6}, 481-493, 1976. 

Mizutani, S., and I. Hattori, Stochastic analysis of bed-thickness
distribution of sediments, {\it Math. Geol., 4}, 123-146, 1972.

Molz, F.J., and G.K. Boman, A fractal-based stochastic interpolation scheme
in subsurface hydrology, {\it Water Resour. Res., 29}, 3769-3774, 1993.

Muller, P., Stochastic forcing of quasi-geostrophic eddies, in {\it Stochastic
Modelling
in Physical Oceanography}, edited by R. J. Adler, P. Muller, and B. Rozovskii,
pp. 381-395,
Birkhauser,
Boston, 1996.

Musha, T., and H. Higuchi, The $1/f$ fluctuation of a traffic current
of an expressway, {\it Jap. J. Appl. Phys., 15}, 1271-1275, 1976. 

Muto, T., The Kolmogorov model of bed thickness distribution: An assessment
based on numerical simulation and field-data analysis, {\it Terra Nova, 7},
408-416, 1995. 

National Geophysical Data Center, {\it Global Paleomagnetic Database, 
version 3.1}, National Oceanographic and Atmospheric Administration, 1995.  

Noakes, D.J., K.W. Hipel, A.I. McLeod, C. Jimenez, and S. Yakowitz,
Forecasting annual geophysical time series, {\it Int. J. Forecasting,
4}, 103-115, 1988.

North, G.R., and R.F. Cahalan, Predicability in a solvable stochastic
climate model, {\it J. Atm. Sci., 38}, 504-513, 1981.

Novikov, E.A., Random force method in turbulence theory, {\it Soviet Phys., J.E.T.P.,
17}, 1449-1454, 1963.

Pal, P.C., and P.H. Roberts, Long-term polarity stability and strength of the
geomagnetic dipole, {\it Nature, 331}, 702-705, 1988.

Palmer, M.W., Fractal geometry: A tool for describing spatial
patterns of plant communities, {\it Vegetatio 75}, 91-102, 1988.

Passier, M.L., and R.K. Snieder, On the presence of intermediate-scale heterogeneity
in the upper mantle, {\it Geophys. J. Int., 123}, 817-837, 1995.

Peixoto, J.P., and A.H. Oort,
{\it Physics of Climate}, Am. Inst. Phys.,
New York, 1992.

Pelletier, J.D., Variations in solar luminosity from time scales of
minutes to months, {\it Astrophys. J., 463}, L41-L45, 1996.

Pelletier, J.D., Analysis and modeling of the natural variability of climate,
{\it J. Climate, 10}, 1331-1342, 1997a.

Pelletier, J.D., Kardar-Parisi-Zhang scaling of the convective boundary
layer and the fractal structure of cumulus cloud fields, {\it Phys. Rev.
Lett., 78}, 2672-2675, 1997b. 

Pelletier, J.D., Why is topography fractal?, 
{\it J. Geophys. Res.}, submitted, 1997c.

Pelletier, J.D., and D.L. Turcotte, Scale-invariant topography and
porosity variations in sedimentary basins, {\it J. Geophys. Res., 101},
28,165-28,175, 1996.

Pilkington, M., and J.P. Todoeschuck, Stochastic inversion for scaling geology,
{\it Geophys. J. Int., 102}, 205-217, 1990.

Platt, T., and K.L. Denman,
Spectral analysis in ecology, {\it Ann. Rev.
Ecol. Syst., 6}, 189-210, 1975.

Pleune, R., Vertical diffusion in the stable atmosphere, {\it Atm.
Env. A, 24}, 2547-2555, 1990.

Plotnick, R.E., A fractal model for the distribution of stratigraphic hiatuses,
{\it J. Geol., 94}, 885-890, 1986.

Plotnick, R.E., and K. Prestegaard, Fractal analysis of geologic
time series, in {\it Fractals in Geography}, edited by L. De Cola and S. Lam,
pp. 193-210,
Prentice-Hall, Englewood Cliffs, Cal., 1993.

Press, W.H., S.A. Teukolsky, W.T. Vetterling, and B.P. Flannery, {\it
Numerical Recipes in C: The Art of Scientific Computing}, 2nd ed., Cambridge
Univ. Press, New York, 1992.

Rosen, J.M., N.T. Kjome, R.T. McKenzie, and J.B. Liley, Decay of
Mount Pinatubo aerosols at midlatitudes in the northern and southern hemispheres,
{\it J. Geophys. Res., 99}, 25,733-25,739, 1994.

Rothman, D.H., J. Grotzinger, and P. Flemings, Scaling in turbidite 
deposition, {\it J. Sed. Petrol. A, 64}, 59-67, 1993.

Sadler, P.M., Sediment accumulation rates and the completeness of
stratigraphic sections, {\it J. Geol., 89}, 569-584, 1981.

Sadler, P.M., and D.J. Strauss, Estimation of completeness of stratigraphical
sections using empirical data and theoretical methods, {\it J. Geol. Soc. London,147}, 471-485, 1990.

Schwarzacher, W., {\it Sedimentation Models and Quantitative Stratigraphy},
Elsevier, Amsterdam, 1975.

Seinfeld, J.H., {\it Atmospheric Chemistry and Physics of Air Pollution},
Wiley, New York, 1986.

Seki, M. and K. Ito, A phase-transition model for geomagnetic polarity
reversals, {\it J. Geomag. Geoelectr., 45}, 79-88, 1993.

Shiomi, K., H. Sato, and M. Ohtake, Broad-band power-law spectra of well-log
data in Japan and the effects on seismic wave propagation, {\it Geophys.
J. Int.}, in press, 1997. 

Slack, J.R., J.M. Landwehr,
Hydro-climatic data network: A U.S. Geological Survey streamflow
data set for the United States for the study of climatic variations:
1874-1988, {\it U.S. Geol. Surv. Open-File Rep. 92-129}, 1992.

Stevenson, D.J., Planetary magnetic fields, {\it Rep. Prog. Phys., 46},
555-620, 1983.

Strauss, D.J., and  P.M. Sadler, Stochastic models for the completeness of
stratigraphic sections, {\it Math. Geol., 21}, 37-59, 1989.

Sugihara, G., and R.M. May, Applications of fractals in ecology,
{\it Trends in Ecol. Evol. 5}, 79-86, 1990.

Takayasu, M., and H. Takayasu, $1/f$ noise in a traffic model, {\it Fractals, 1},
860-866, 1993.

Tarboton, D.G., The source hydrology of severe sustained drought in the
Southwestern United States, {\it J. Hyd., 161}, 31-69, 1994.

Tauxe, L., and P. Hartl, 11 million years of Oligocene geomagnetic
field behavior, {\it Geophys. J. Int., 128}, 217-229, 1997.    

Thibal, J., J.-P. Pozzi, V. Barthes, and G. Dubuisson, Continuous record
of geomagnetic field intensity between 4.7 and 2.7 Ma from downhole measurements,
{\it Earth Plan. Sci. Lett., 136}, 541-550, 1995.

Thouveny, N., K.M. Creer, and I. Blunk, Extension of the Lac du
Bouchet palaeomagnetic record over the last 120,000 years, {\it Earth Plan.
Sci. Lett., 97}, 140-161, 1990.

Tipper, J.C., Rates of sedimentation, and stratigraphical 
completeness, {\it Nature,
21}, 296-298, 1983.

Tjemkes, S.A., and M. Visser, Horizontal variability of temperature,
specific humidity, and cloud liquid water as derived from spaceborne
observations, {\it J. Geophys. Res., 99}, 23,089-23,105, 1994.

Todoeschuck, J.P., O.G. Jensen, and S. Labonte, Gaussian scaling noise 
model of seismic reflection sequences: evidence from well logs, {\it Geophysics,
55}, 480-484, 1990.

Tubman, K.M., and S.D. Crane, Vertical versus horizontal well log
variability and application to fractal
reservoir modeling, in
{\it Fractals in Petroleum Geology and Earth Sciences},
edited by C.C. Barton and P.R. LaPointe, pp. 279-294, Plenum,
New York, 1995.

Turcotte, D.L., A fractal interpretation of topography and geoid spectra on the
Earth,
Moon, Venus, and Mars, {\it J. Geophys. Res., 92}, 597-601, 1987.

Turcotte, D.L., {\it Fractals and Chaos in Geology and Geophysics},
Cambridge Univ. Press, New York, 1992. 

Turner, G.M., and R. Thompson, Lake sediment record of the geomagnetic 
secular variation in Britain during Holocene times, {\it Geophys. J. R.
Astron. Soc., 65}, 703-725, 1981. 

Valet, J.-P., L. Meynadier, Geomagnetic field intensity and reversals
during the past four million years, {\it Nature, 366}, 234-238, 1993.

van der Ziel, A., On the noise spectra of semiconductor noise and of flicker
effect, {\it Physica, 16}, 359-375, 1950.

Van Kampen, N.G., {\it Stochastic Processes in Physics and Chemistry},
North-Holland, Amsterdam, 1981.

Van Vliet, K.M., A. van der Ziel, and
R.R. Schmidt, Temperature-fluctuation
noise of thin films supported by a substrate, {\it J. Appl. Phys., 51},
2947-2956, 1980.

Verosub, K.L., P.J. Mehringer, and P. Waterstraat, Holocene secular 
variation in western North America: paleomagnetic record from Fish Lake,
Harney County, Oregon, {\it J. Geophys. Res., 91}, 3609-3623, 1986.

Vicsek, T., {\it Fractal Growth Phenomena}, World Sci., River Edge, N. J., 1992.

Vistelius, A.B., and T.S. Feigel'son, On the theory of bed formation, {\it Dokl.
Akad. Nauk. SSSR, 164}, 158-160, 1965.

Voorhies, C.V., and J. Conrad, Accurate predictions of mean geomagnetic
dipole excursion and reversal frequencies, mean paleomagnetic field intensity,
and the radius of the Earth's core using McLeod's rule, {\it NASA Technical
Memorandum 104634}, 1996.

Vose, R.S., R.L. Schmoyer, P.M. Stewer, T.C. Peterson, R. Heim, T.R. Karl,
and J.K. Eischeid, The global historical climatology network: Long-term
monthly temperature, precipitation, sea-level pressure, and station pressure
data, {\it Envir. Sci. Div. Pub. No. 392}, Oak Ridge National
Laboratory, 1992.

Voss, R.F., and J. Clarke, Flicker ($1/f$) noise: Equilibrium temperature
and resistance fluctuations, {\it Phys. Rev. B, 13}, 556-573, 1976.

Walden, A.T., and J.W.J. Hosken, An investigation of the spectral properties of
primary reflection coefficients, {\it Geophys. Prospect., 33}, 400-435, 1985.

Wallis, J.R., D.P. Lettenmaier, and E.F. Wood, A daily hydroclimatological
data set for the continental U.S., {\it Water Resour. Res., 27},
1657-1663, 1991.

Weissman, M.B., $1/f$ noise and other slow, nonexponential kinetics in condensed
matter, {\it Rev. Mod. Phys., 60}, 537-571, 1988.

Wilde, P., W.R. Normark, and T.E. Chase, Channel sands and petroleum potential
of Monterey deep-sea fan, California, {\it AAPG Bull., 62},
967-983, 1978.

Wunsch, C., and D. Stammer, The global frequency-wavenumber spectrum of oceanic
variability estimated from TOPEX/POSEIDON altimetric measurements, {\it J. Geophys.
Res., 100}, 24,895-24,910, 1995.

Yiou, P., J. Jouzel, S. Johnsen, and O.E. Rognvaldsson, Rapid oscillations
in Vostok and GRIP ice cores, {\it Geophys. Res. Lett., 22}, 2179-2182, 1995.  

\newpage

\section*{Figure Captions}

Figure 2.1: Atmospheric temperatures at Vostok, Antarctica inferred
from Deuterium concentrations in the Vostok ice core [{\it Jouzel et al.}, 1987].   

Figure 2.2: Power-spectral density estimated with the 
Lomb periodogram of the temperature inferred from the
Deuterium concentrations in the Vostok (East Antarctica) ice core.
The power-spectral density $S$ is given as a function of frequency for time scales 
of 500 yr to 200 kyr.

Figure 2.3: Average monthly atmopsheric temperature for Central England 
[{\it Manley} [1974] with the yearly periodicity removed. 

Figure 2.4: Power-spectral density of the time series of Central England
temperatures in Figure 2.3. 

Figure 2.5: Average power-spectral density of 94 complete monthly 
temperature time series from the dataset of the {\it Vose et al.} [1992]
plotted as a function of frequency in yr$^{-1}$. The power-spectral density
$S$ is given as a function of frequency for time scales of
2 months to 100 yr.     

Figure 2.6: Average power-spectral density of 50 continental
daily temperature time series from the data set of the 
{\it National Climatic Data Center} [1994] as a function of frequency
in yr$^{-1}$. The power-spectral density $S$ is given as a function 
of frequency for time scales of 2 days to 10 yr. 

Figure 2.7: Average power-spectral density of 
50 maritime daily temperature time series from the data set of the
{\it National Climatic Data Center} [1994] as a function of frequency
in yr$^{-1}$. The power-spectral density $S$ is given as a function
of frequency for time scales of 2 days to 10 yr.

Figure 2.8:
Power-spectral density of local atmospheric temperature from instrumental
data and inferred from ice cores from time scales of 200 kyr to 2 days. The
high frequency data are for continental stations. Piecewise power-law
trends are indicated.

Figure 2.9: Average normalized power-spectral density of 636 monthly river discharge
series as a function of frequency in yr$^{-1}$. 

Figure 2.10: Average normalized power-spectral density of 43 tree ring chronologies
in the western U.S. as a function of frequency in yr$^{-1}$.

Figure 2.11: Average power-spectral density of the number of random
walkers in the central site of a lattice. The average of 50 simulations
is presented.

Figure 2.12: Cumulative distribution function of the time series produced
by the stochastic diffusion model (solid circles). The curve
represents the cumulative log-normal distribution function fit to the
data.

Figure 2.13: (a) Geometry of the one-dimensional diffusion 
calculation detailed in the text. (b) Boundary conditions appropriate
to the air masses above the ocean (maritime stations), where the ocean acts
as a thermal conductor. (c) Boundary conditions appropriate to the air
masses above the continents (continental stations), where the continents 
act as a thermal insulator.

Figure 2.14: Geometry of the coupled atmosphere-ocean model and the constants
chosen.

Figure 2.15: Average power-spectral density of atmospheric temperature above 
continents and oceans for each grid point in the general circulation model
calculations of {\it Manabe and Stouffer} [1996]. The straight-line 
corresponding to $f^{-\frac{1}{2}}$ is included for comparison. 

Figure 2.16: Power spectral density of variations in the solar
irradiance in 1987 and 1985 from the ACRIM project as a function of 
frequency in hour$^{-1}$.

Figure 2.17: Logarithm (base 10) of the recurrence interval as a function
of drought duration and magnitude (normalized to the mean annual flow)
for a log-normal distribution with coefficient of variation (a) 0.2,
(b) 0.4, and (c) 0.6.

Figure 3.1: Illustration of the sediment deposition model. In each case a
site is chosen randomly (the center of the three sites in each of the
above pictures).
The dashed block is the
unit of sediment being added to the surface. The arrows point toward
the site upon which the unit of sediment will be deposited.
(a) The chosen site has a lower
elevation than either of its nearest neighbors, so the sediment is deposited
at the chosen site. (b) One of the nearest neighboring sites has a
lower elevation and the sediment is deposited at that lower site. (c) In the
case of a tie for the lowest elevation between
two or three sites, the site on which the sediment
is deposited is chosen randomly between the sites of the same elevation.

Figure 3.2: A typical surface produced by the deposition model with
1024 grid points.
 
Figure 3.3: Average power spectrum of the surfaces
constructed from fifty independent simulations on 1024 grid points 
as a function of the wave number $k$. The model surfaces are Brownian walks. 

Figure 3.4: One-dimensional 
transect of hillslope topography perpendicular to the channel
dip.
Obtained with the use of laser altimetry [{\it Dunne, Whipple, and Aubry}, 1995].

Figure 3.5: Difference from the mean height of the central site of the lattice
as a function of time.
 
Figure 3.6: Average power spectrum of the difference from the
mean height of the central site for fifty independent simulations as
as a function of frequency $f$. The
power spectrum is proportional to $f^{-\frac{3}{2}}$.

Figure 3.7: Vertical porosity well log from the Gulf of Mexico.

Figure 3.8:
Power spectral density of porosity as a function of
wave number in units of m$^{-1}$ in fifteen wells
from the
Gulf of Mexico. The spectra are offset so that they may be placed on the
same graph.

Figure 3.9: Probability density function for elevation of topographic transects
from
[{\it Dunne et al.}, 1995].

Figure 3.10: Wells producing hydrocarbons in the (a) Powder River and (b)
Denver basins [{\it Barton and Scholz}, 1995]. 
Distance units are scaled
such that the basin is 128 x 128.
 
Figure 3.11: Pair-correlation function of the Powder River
and Denver basins as a function of the pair
separation.
 
Figure 3.12: Synthetic reservoir constructed from a
source and cap rock with a two-dimensional Brownian walk topography
constructed on a 128 x 128 grid where all the sites with porosity greater than
a fixed level are showing.
 
Figure 3.13: Pair-correlation functions for synthetic
reservoirs with caprocks constructed with different values of $\beta$. 
The plots are
offset so that they may be placed on the same graph.

Figure 3.14: 
The nondimensional thickness of sediments $h\sigma /D$ in a sedimentary
basin is given as a function of nondimensional time $t\sigma ^{2} /D$ for a
sequence in which the ratio of the standard deviation to the mean of
sedimentation,
$\sigma /\overline{\eta}$
is 0.1. 

Figure 3.15: For the model given in Figure 3.14 the age of the sediments is given
as a function of depth. Only those sediments which are not later eroded
are preserved.

Figure 3.16: Illustration of a model for sediment deposition based on a Devil's
staircase associated with a second-order Cantor set. (a) Age of sediments
$T$ as a function of depth $y$. (b) Illustration of how the Cantor set
is used to construct the sedimentary pile. (c) Average rate of deposition
$R$ as a function of the period $T$ considered.

Figure 3.17: Average rate of sedimentation, $R/ \overline{\eta}$, as as
function of time span, $T\sigma ^{2}/D$,
for the sediment column of Figure 3.10(b).

Figure 3.18: Observed sedimentation rates as a function of time span
from the data of {\it Sadler} [1995]. The data have been binned
in equally-spaced bins in log space.
A least-square
linear fit to the logarithms of the data yields a slope of $-\frac{3}{4}$
indicating that $R\propto T^{-\frac{3}{4}}$.

Figure 3.19: Cumulative frequency-length distribution of hiatuses, the number of
hiatuses longer than nondimensional hiatus length
$t_{h}\sigma ^{2}/D$, for synthetic sequences
produced with the stochastic diffusion model.

Figure 3.20: Cumulative frequency-thickness distribution of nondimensional
bed thicknesses
for synthetic sediment columns with $\sigma /\overline{\eta}
=0.1,0.01,0.001$ and $0.0003$. The distributions are exponential.
 
Figure 3.21: Cumulative frequency-thickness distribution of bed thicknesses of 
deep-sea sequences from
(a) Ra Stua, (b) Castagne, and (c) Cismon Valley, Italy published in
{\it Claps and Masetti} [1994]. 
The coefficients in the exponential distributions determined
by a least-squares fit of the logarithm of the bed number to the bed thickness for the largest
forty beds were -0.052, -0.166, and -0.252, showing an increasing trend with sedimentation
rate consistent with the model behavior.
 
Figure 4.1: 
Paleointensity of the virtual axial dipole moment (VADM) of the
earth's magnetic field (with reversed polarity data
given by negative values)
inferred from sediment cores for the past 4 Ma from {\it Meynadier} [1994].

Figure 4.2: Power spectral density 
of the geomagnetic field intensity variations estimated
from the Lomb periodogram of sediment
cores from {\it Meynadier} [1992] and {\it Meynadier} [1994] 
and archeomagnetic data from
{\it Kovacheva} [1980]. The power-spectral density $S$ is given as a function of
frequency $f$ for time scales of 100 yr to 4 Myr.

Figure 4.3: Cumulative 
frequency-length distribution of the lengths of polarity intervals from the
time scale of {\it Harland et al.} [1990] (solid curve), {\it Cande and
Kent} [1992a,1995] (dashed curve), and the {\it Cande and Kent} [1992a,1995]
time scale from C1 to C13 with cryptochrons included (dashed-dotted line).

Figure 4.4: 
A $1/f$ noise with a normal distribution with mean of 8.9 and standard
deviation of 3.4x10$^{22}$Am$^{2}$ representing the geomagnetic field intensity in
one polarity state.  

Figure 4.5:
Binormal $1/f$ noise constructed from the normal $1/f$ noise of Figure 4.4
as described in the text.

Figure 4.6: Cumulative frequency-length polarity interval distributions from the
{\it Harland et al.} [1990] time scale and that of the binormal $1/f$ noise model of
intensity variations. The distribution from the {\it Harland et al.} [1990] time scale
is the dashed curve. The solid line is the
average cumulative distribution from the $1/f$ noise model. The dotted lines
represent the minimum and maximum reversal length distributions for 20 numerical
experiments, thereby representing 95\% confidence intervals. 
 
Figure 4.7: Cumulative frequency-length polarity interval distributions for the
$1/f$ noise model of intensity variations (shown in the middle, the same as that
in Figure 4.6) and for intensity variations with power spectra proportional
to $f^{-0.8}$ and $f^{-1.2}$. This plot
illustrates that the polarity length distribution is very sensitive to the form
of the power spectrum, allowing us to conclude that the agreement between the
model and the observed distribution in Figure 4.5 is unique to $1/f$ noise intensity
variations.
 
Figure 4.8: Pair-correlation function of the reversal history according to the
Harland et al. (1990) time scale (filled circles), Cande and Kent (1992a,1995) (unfilled circles), synthetic reversals produced from $1/f$ noise
model of intensity variations (boxes), and a Poisson process (triangles).
The real and synthetic reversals exhibit similar non-random clustering.

Figure 4.9: Magnetic field inclination inferred from the Lac du Bouchet
sediment core [{\it Thouveny et al.}, 1990].
 
Figure 4.10: Power spectra of inclination and declination from the Lac du
Bouchet sediment core. The declination spectrum is offset from the inclination
spectrum so that they may be placed on the same graph.

Figure 4.11: Power spectra of inclination from the following locations,
top to bottom: 1) Anderson Pond, 2) Bessette Creek, 3) Fish Lake, 4) 
Lake Bullenmerri, and 5) Lake Keilambete. The spectra are offset to place
them on the same graph.

Figure 4.12: 
Dipole moment produced by the model for geomagnetic variations
normalized to the average dipole moment, set to be one.
The field exhibits reversals
with a broad distribution of polarity interval lengths and a variable reversal
rate decreasing at later times in the simulation.

Figure 4.13: Distribution of the magnetic field according to ten simulations of
the model (solid curve)
and a binormal distribution fit to the data (dashed line). 
The binormal distribution fits
the data well. 

Figure 4.14: Average power spectrum of the mean value of the magnetic field
(dipole field) from 25 simulations. The spectrum has a low-frequency spectrum
with $S(f)\propto f^{-1}$ and a high frequency region $S(f)\propto f^{-2}$.
The same spectrum is observed in geomagnetic intensity from sediment cores
and historical data.  

Figure 4.15: Average power spectrum of the angular deviation from a dipole
field from 25 simulations. The spectrum is $S(f)\propto f^{-2}$ for high
frequencies and gradually flattens out to a constant spectrum at low
frequencies.   
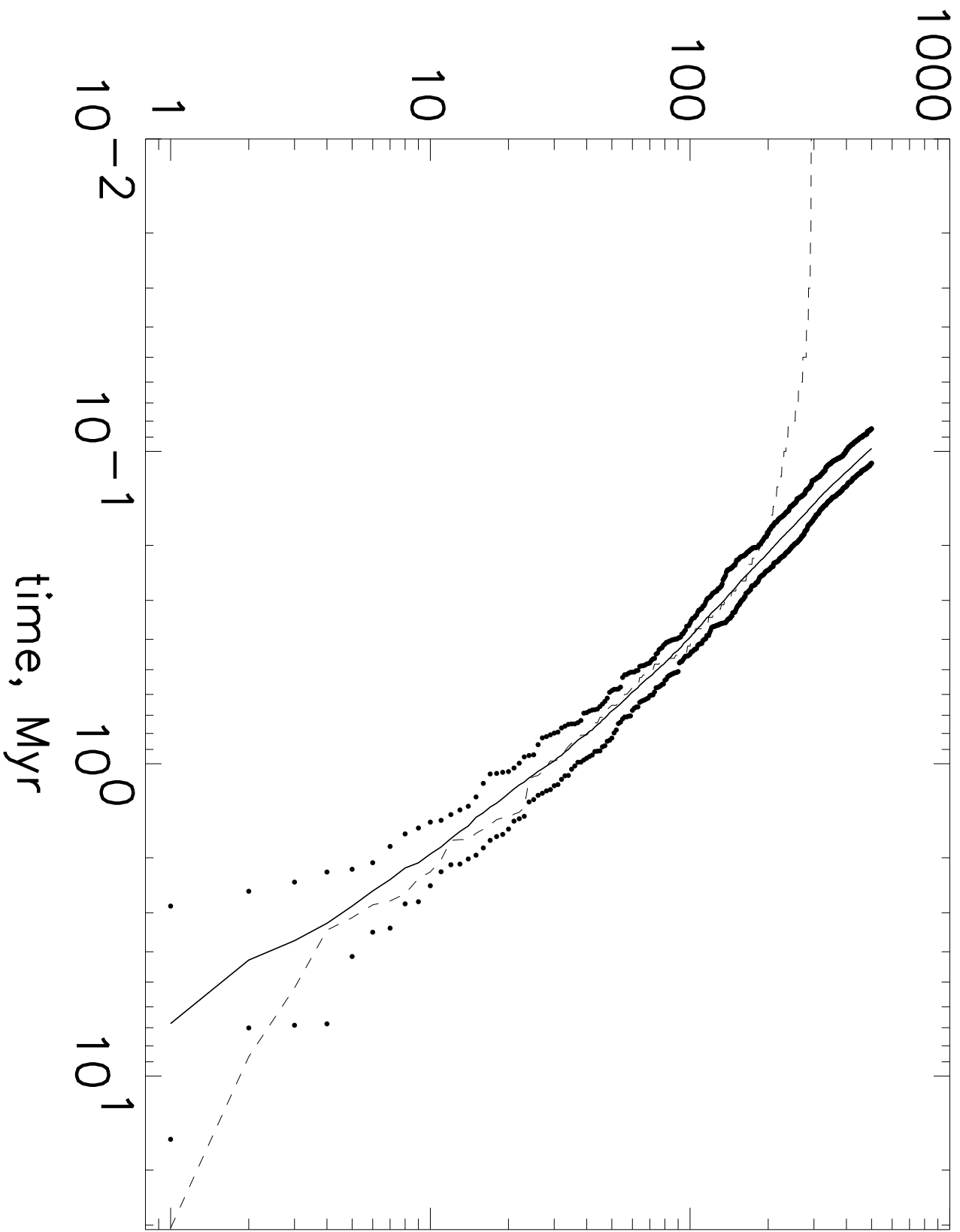
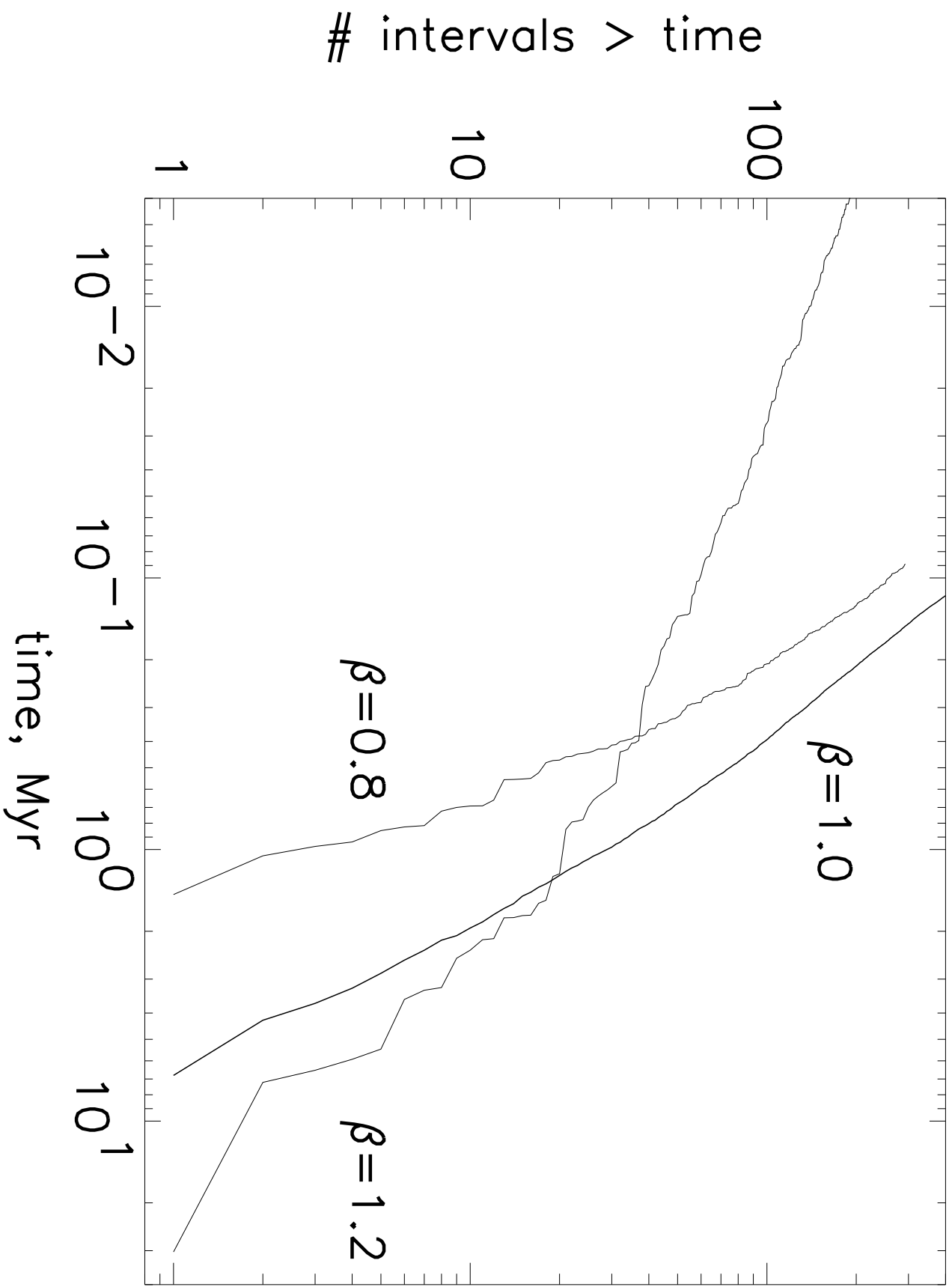
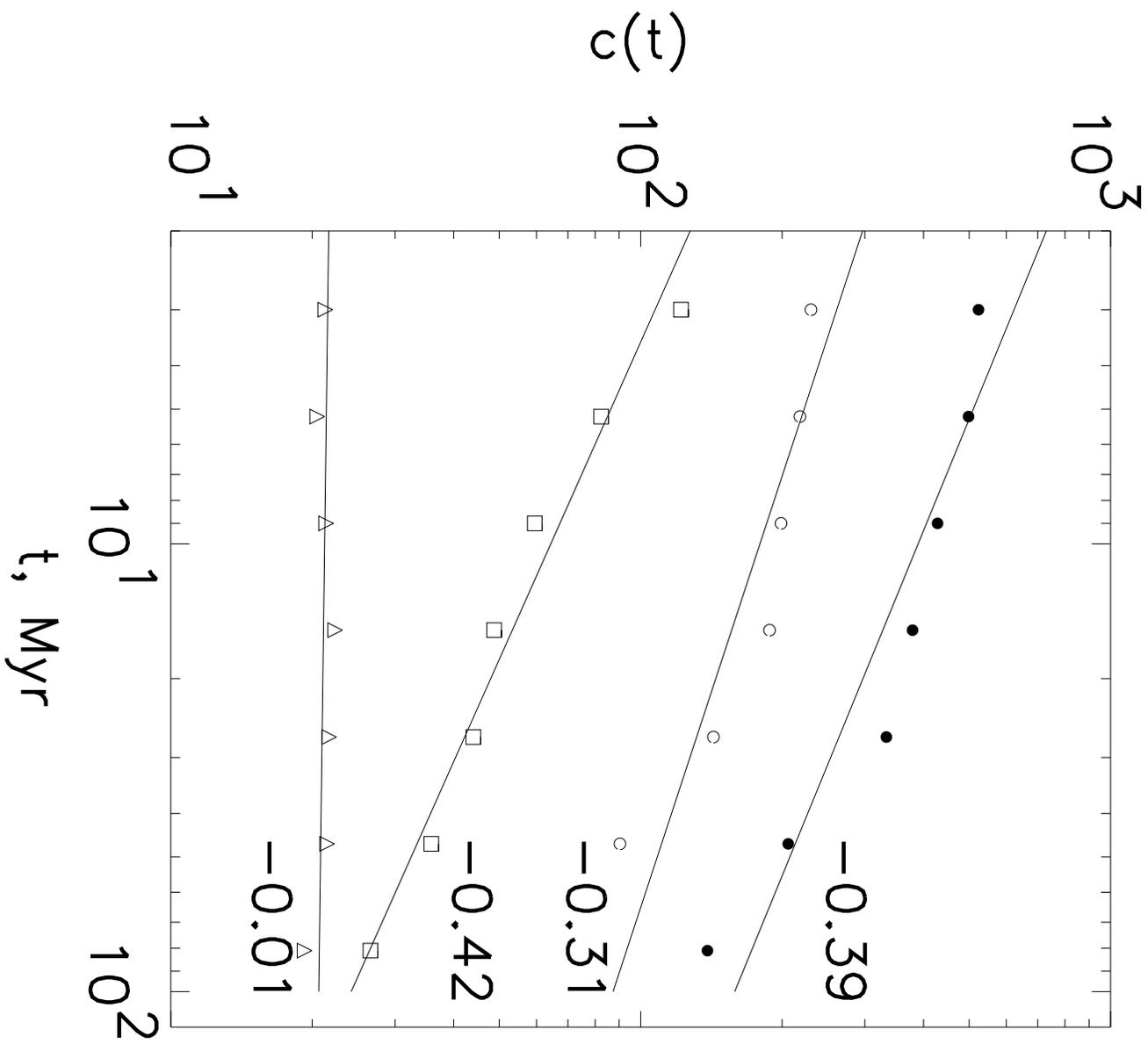
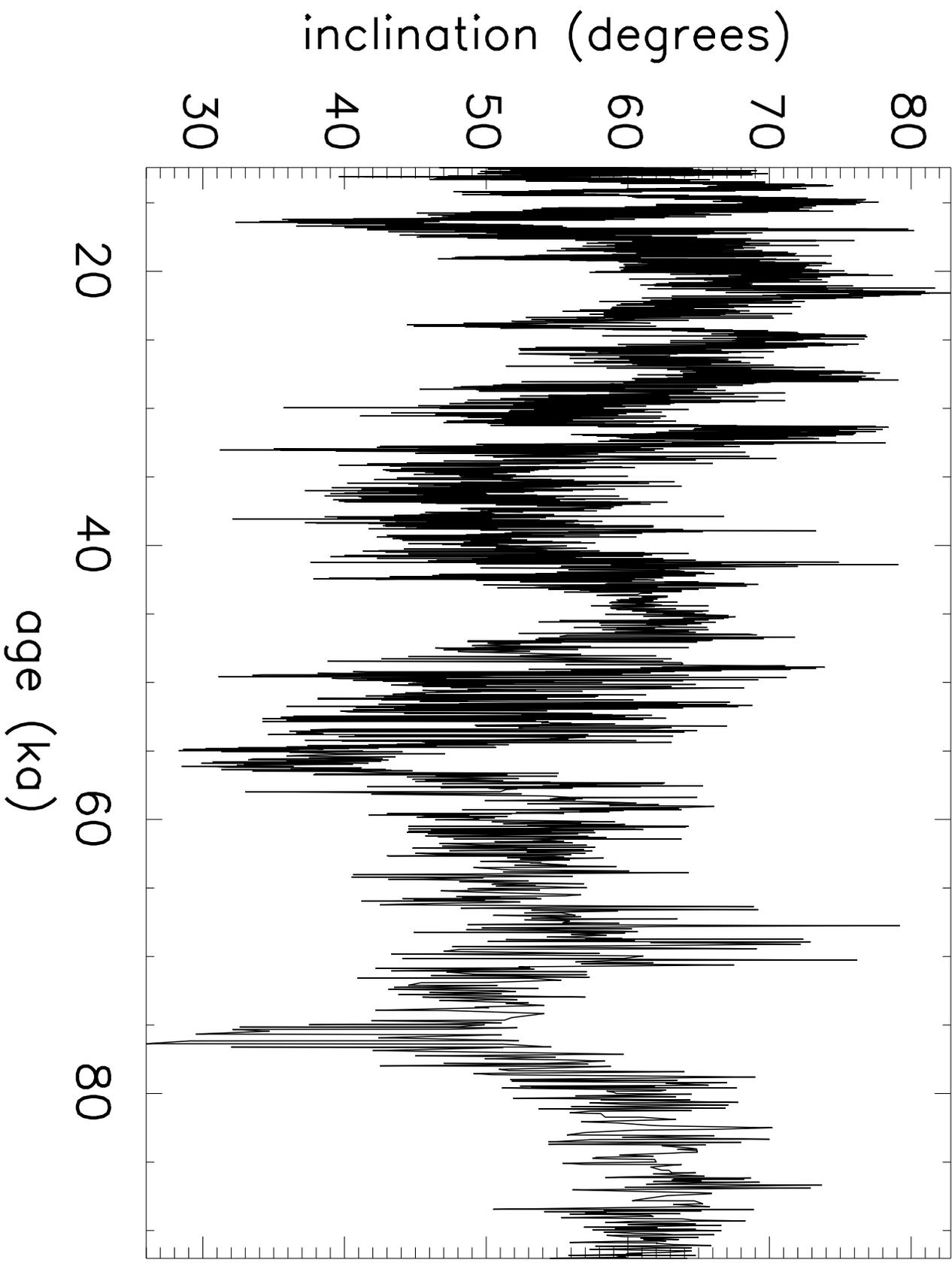
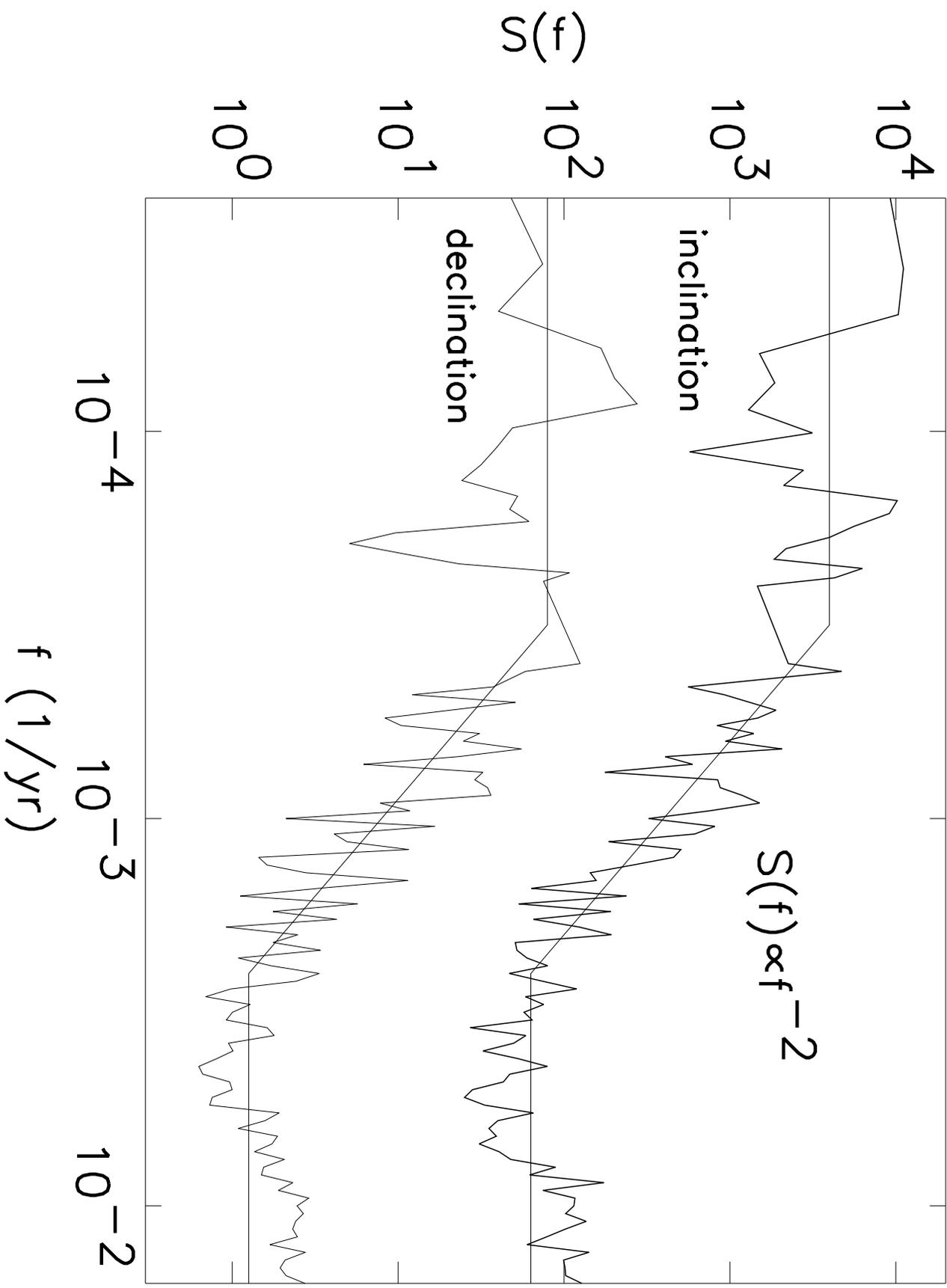
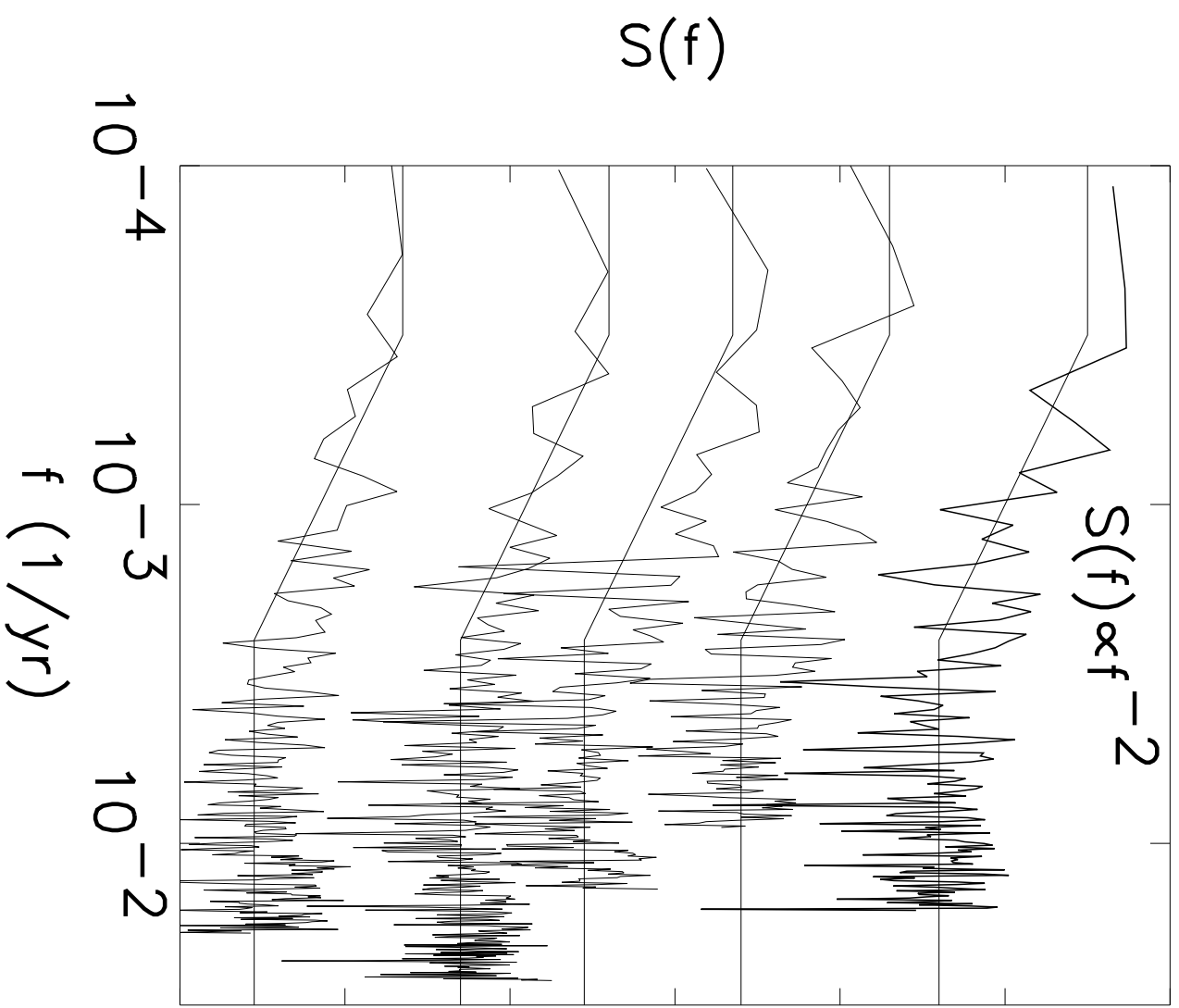
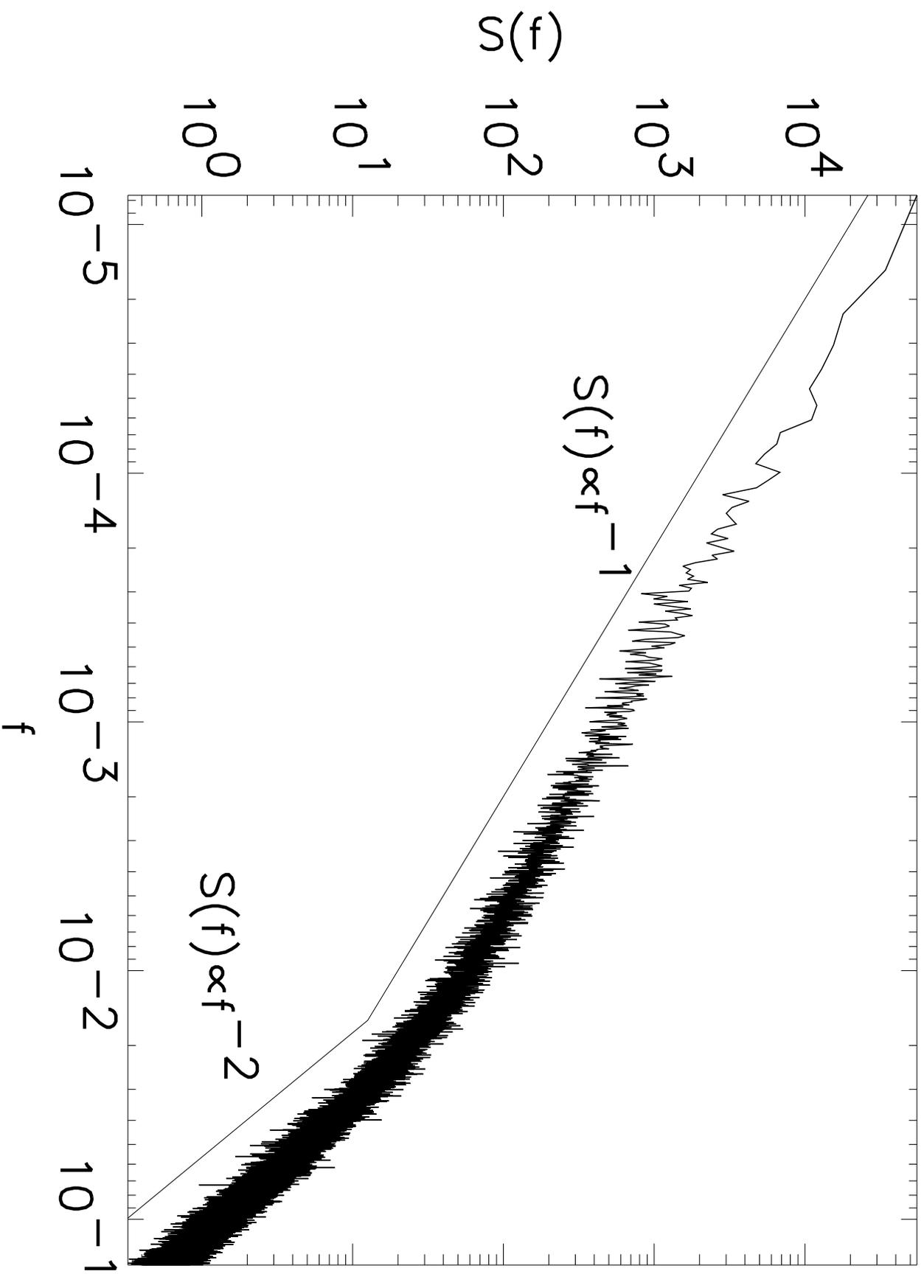
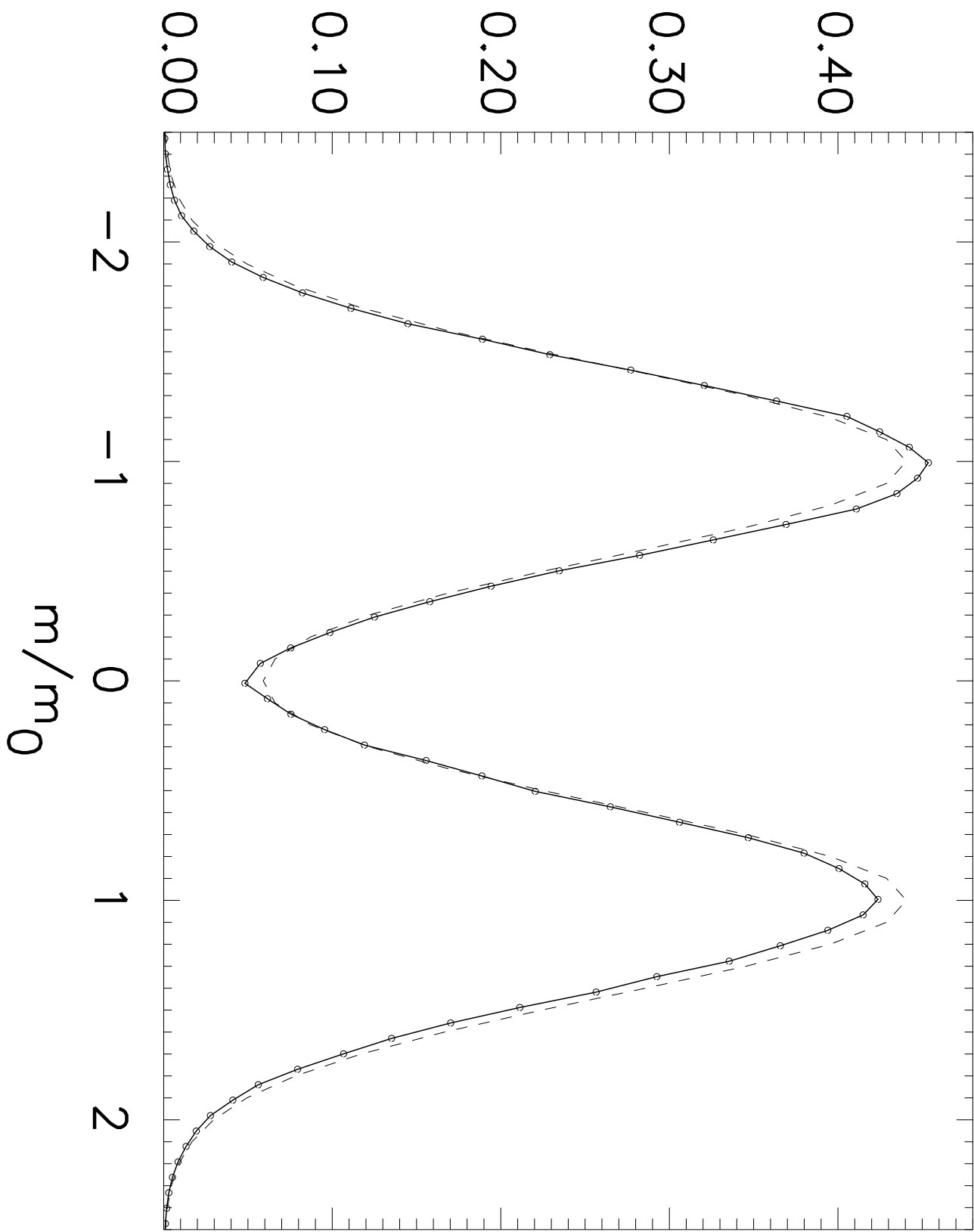
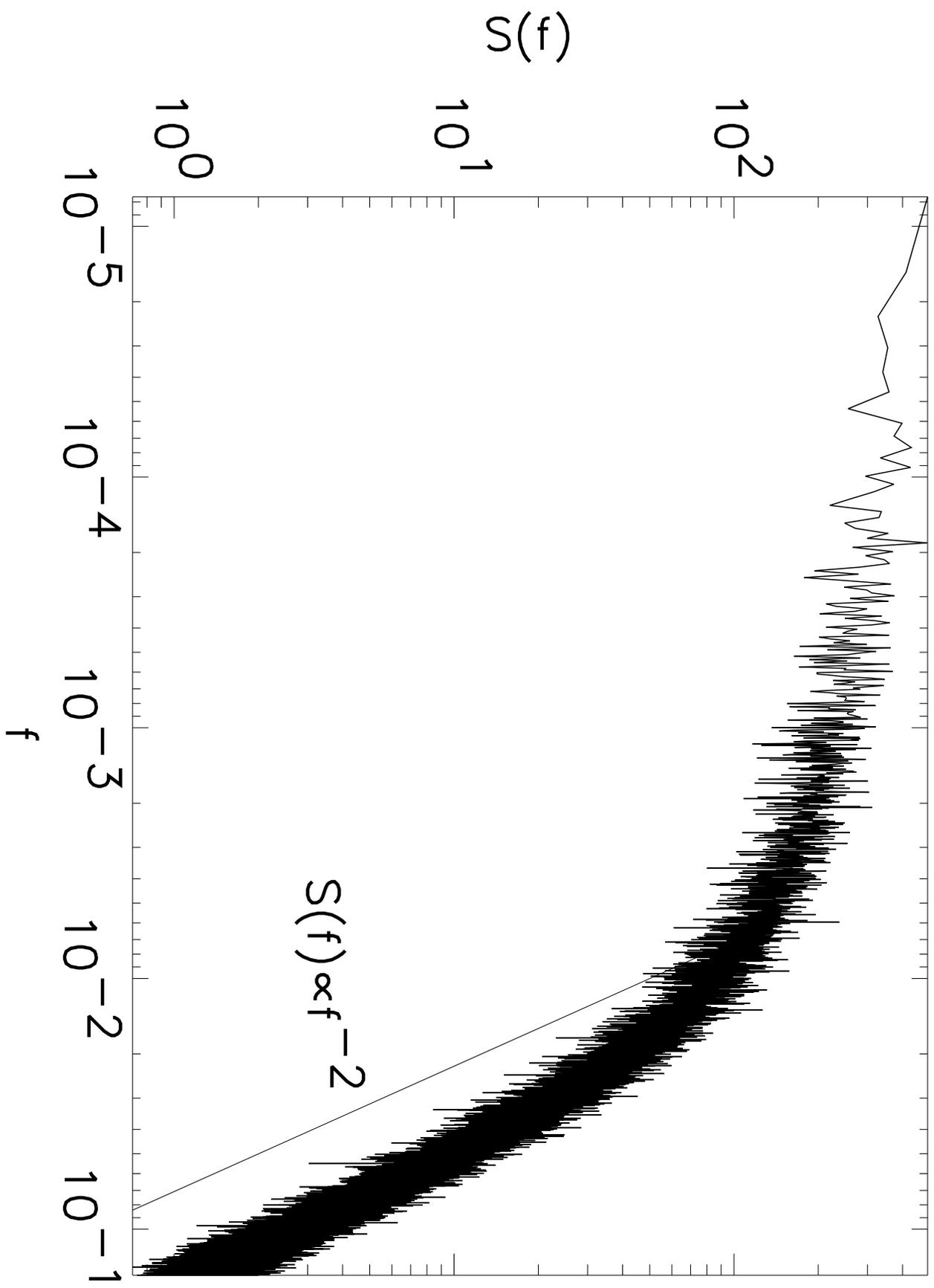
\end{document}